\documentclass[twocolumn,aps,showpacs,superscriptaddress,floatfix]
{revtex4}
\usepackage[dvips]{graphicx}
\newcommand{\ket}[1]{\left | #1 \right \rangle}
\newcommand{\bra}[1]{\left \langle #1 \right |}
\newcommand{\proj}[1]{\ket{#1} \bra{#1}}
\newcommand{\varket}[1]{\left | #1 \right )}
\newcommand{\varbra}[1]{\left ( #1 \right |}

\newcommand{\tr}{{\rm \, Tr }\, }
\newcommand{\bm}[1]{\mbox{\boldmath{$#1$}}}

\newcommand{\beq}{\begin{equation}}
\newcommand{\eeq}{\end{equation}}
\newcommand{\beqa}{\begin{eqnarray}}
\newcommand{\eeqa}{\end{eqnarray}}
\newcommand{\beqan}{\begin{eqnarray*}}
\newcommand{\eeqan}{\end{eqnarray*}}

\begin{document}
\title{Multimode theory of measurement-induced 
non-Gaussian operation on wideband squeezed light}
\author{Masahide Sasaki}
\email{psasaki@nict.go.jp}
\affiliation{
    National Institute of Information and Communications 
    Technology, 
    4-2-1 Nukui-Kita, Koganei, Tokyo 184-8795, Japan}
\affiliation{
    CREST, Japan Science and Technology Agency, 
    1-9-9 Yaesu, Chuoh, Tokyo 103-0028, Japan}
\author{Shigenari Suzuki}
\affiliation{
    National Institute of Information and Communications 
    Technology, 
    4-2-1 Nukui-Kita, Koganei, Tokyo 184-8795, Japan}
\affiliation{
    Department of Electronics and Electrical Engineering,
    Keio University, \\
    3-14-1 Hiyoshi Kohoku, Yokohama 223-8522, Japan}
%
%\author{Kentaro Wakui}
%\affiliation{
%    National Institute of Information and Communications 
%    Technology, 
%    4-2-1 Nukui-Kita, Koganei, Tokyo 184-8795, Japan}
%\affiliation{
%    Department of Applied Physics, The University of Tokyo 
%    7-3-1 Hongo, Bunkyo-ku, Tokyo 113-8656, Japan}
%\affiliation{
%    CREST, Japan Science and Technology Agency, 
%    1-9-9 Yaesu, Chuoh, Tokyo 103-0028, Japan}
%
%\author{Kenji Tsujino}
%\affiliation{
%    National Institute of Information and Communications 
%    Technology, 
%    4-2-1 Nukui-Kita, Koganei, Tokyo 184-8795, Japan}
%\affiliation{
%    CREST, Japan Science and Technology Agency, 
%    1-9-9 Yaesu, Chuoh, Tokyo 103-0028, Japan}
%
%\date{\today} 
%
%
%
\begin{abstract}
We present a multimode theory of non-Gaussian operation 
induced by an imperfect on/off-type photon detector 
on a splitted beam from a wideband squeezed light. 
The events are defined for finite time duration $T$ 
in the time domain. 
The non-Gaussian output state is measured by the homodyne detector 
with finite bandwidh $B$. 
Under this time- and band-limitation to the quantm states, 
we develop a formalism to evaluate the frequency mode matching 
between 
the on/off trigger channel 
and 
the conditional signal beam in the homodyne channel. 
Our formalism is applied to the CW and pulsed schemes. 
We explicitly calculate the Wigner function of the conditional 
non-Gaussian output state in a realistic situation. 
Good mode matching is achieved for $BT\alt1$, 
where the discreteness of modes becomes prominant, 
and only a few modes become dominant 
both in the on/off and the homodyne channels. 
If the trigger beam is projected nearly onto 
the single photon state in the most dominant mode 
in this regime, 
the most striking non-classical effect will be observed 
in the homodyne statistics. 
The increase of $BT$ and the dark counts 
degrades the non-classical effect. 
In the CW scheme, 
one will be able to attain a stringent mode matching 
in the regime $BT\alt0.1$.  
Actually the bandwidth can be set typically to $B\approx10$\, MHz 
by using an appropriate set of filters, 
while the duration $T$ can be set in 10\,ns order 
by electrical gating. 
The spatial mode matching can be fulfilled 
by careful cavity locking. 
The temporal mode matching is not a problem in that time scale. 
So the CW scheme will provide a good test-bed for non-Gaussian 
operations. 
In the pulsed scheme, 
the duration $T$ is automatically set 
by the laser pulse width, which is typically of ps order. 
The band limitation comes from the LO bandwidth 
$\approx1$\,THz, 
corresponding to the Fourier-transform limit $BT\approx1$. 
This still satisfies the condition 
for observing the non-classical effect 
provided that the mode matching in terms of the other degrees 
of freedom is perfect. 
In practice, however, it may be more challenging to realize 
a high quality spatiotemporal mode matching in the pulsed scheme 
than the CW setting. 
\end{abstract}
%
% 42.50.Dv Nonclassical states of the electromagnetic field, 
%          including entangled photon states; 
%          quantum state engineering and measurements 
% 03.65.Wj State reconstruction, quantum tomography 
% 03.67.Mn Quantum entanglement production, characterization, 
%          and manipulation 
% 03.67.Pp Quantum error correction and other methods for 
%          protection against decoherence 
% 42.50.Md Optical transient phenomena: 
%          quantum beats, photon echo, free-induction decay, 
%          dephasings and revivals, optical nutation, 
%          and self-induced transparency 
% 78.20.Bh Optical properties, condensed matter spectroscopy 
%          and other interaction of radiation and particles with 
%          condensed matter: 
%          Theory, models, and numerical simulation
%
%\date{\today}
\maketitle

%%%%%%%%%%%%%%%%%%%%%%%%%%%%%%%%%%%%%%%%%%%%%%%%%%%%%%%%%%%%%%%
\section{Introduction}
\label{Introduction}
%%%%%%%%%%%%%%%%%%%%%%%%%%%%%%%%%%%%%%%%%%%%%%%%%%%%%%%%%%%%%%%
Homodyning and photon counting are standard techniques 
to measure quantum states of optical fields. 
The former accesses the continuous nature of optical fields, 
while the latter does the discrete nature of them.  
The homodyning technique can now have a quantum efficiency 
exceeding 99\% at certain wavelengths, 
and has been successfully applied to various tasks 
in quantum optics and quantum information science. 
In particular, 
it provides a powerful tool to completely characterize 
a quantum state by reconstructing the Wigner function of 
the density operator, 
referred to as the quantum state tomography.

Photon counting technique, on the other hand, 
still fails in attaining near-unit quantum efficiency and 
single photon resolution. 
These properties, however, are not always necessary. 
In fact the on/off type photon counting technique 
with less than unit quantum efficiency 
can be useful for certain applications.  
One of such applications is 
the conditional photon substraction from the squeezed light. 
That is, 
a small fraction of the squeezed light is beamsplitted 
as trigger photons, and guided into the on/off detector. 
By making the amount of fraction of the trigger beam
small enough, 
the detection click effectively realizes the single photon 
subtraction from the main signal beam. 
This method allows one to conditionally generate 
the Schr\"{o}dinger-cat-like state 
\cite{Dakna97}, 
and also to conditionally increase the entanglement of 
the bipartite squeezed beams 
\cite{Opatrny00,Cochrane02,Browne03}. 
More importantly 
these conditional operations are non-Gaussian operations 
that can be implemented with current technologies. 
The importance of non-Gaussian operations on continuous variables 
(CVs) 
is highlighted by the following no-go statements 
concerning the Gaussian operations. 
Firstly, 
quantum speed up is impossible for harmonic oscillators 
by Gaussian operations with Gaussian inputs 
\cite{Bartlett02PRLs}. 
Secondly, 
the distillation of Gaussian entanglement from 
two Gaussian entangled states is impossible 
only by Gaussian LOCC based on homodyne detection 
\cite{Eisert02,Fiurasek02,Giedke02}. 
Therefore, implementation of non-Gaussian operations is crucial 
to extract ultimate potential of photonic quantum information 
processing.

The first experimental demonstration of 
the measurement-induced non-Gaussian operation 
was done by Wenger et al. in the ultra-short pulsed regime 
\cite{Wenger04}. 
They could observe the phase sensitive non-Gaussian distributions, 
that is, a small dip around the origin of the phase space 
in an unisotropic Wigner function distribution. 
In order to observe the negative dip in the Wigner function 
distribution, which is a strong indication of the non-classicality 
of the state, more careful mode matching considerations should 
be taken.

Ideally 
the main signal beam and the trigger beam must be prepared 
in the same mode. 
That is, trigger photons detected by the photon counter 
must be in a mode obseved by the homodyne detector 
with respect to the spatial, temporal, frequency and 
polarization modes. 
Otherwise the homodyne detector will see the modes 
that are not quantum correlated with the trigger photons, 
degrading the non-Gauusian operations significantly.

The mode matching problem in this kind of conditional operation 
was first studied by Ou 
\cite{Ou97QSemiclOpt}. 
He studied the mode matching 
in the scheme of homodyne measurement of 
a conditionally prepared single photon state 
by measurements on a biphoton state produced in 
the parametric down-conversion. 
This scheme was originally proposed by Yurke and Stoler 
\cite{Yurke_Stoler87}, 
which was analyzed in a single mode 
assuming the perfect mode matching. 
Ou studied it in more practical situations, and 
showed that one has to use a narrow spectral filter 
in the trigger channel 
in order to match the conditionally prepared single photon mode 
to the one of the LO.

Grosshans and Grangier extended the analysis, 
and provided a useful formula 
to renormalize the time and frequency overlap 
between the signal and trigger wave packets 
into an effective quantum efficiency of the homodyne detector 
\cite{Grosshans01}. 
They studied the two cases: 
(i) continuous experiment where the pump and the LO beams are 
initially continuous wave (CW) with small enough linewidths, 
and
(ii) pulsed experiment where pump pulses short enough are used 
so that their linewidth becomes greater than 
the frequency filter bandwidth for the trigger photons. 
They concluded that the mode matching condition is much more 
easily fulfilled in the pulsed regime.

The mode matching in the pulsed regime was further studied 
by Aichele et al. 
including more general models for 
the spatial and spectral filters in the trigger channel 
\cite{Aichele Lvovsky Schiller 02EurPhysJD}. 
They performed an explicit calculation 
of the degree of mode matching in terms of both 
frequency-momentum and 
time-space 
representations, 
and showed ways to attain the optimal mode matching 
by using narrowband filters in the trigger channel.  
Viciani et al. further investigated temporal and spectral 
properties of entangled photon pair for various filter functions 
\cite{Viciani Zavatta Bellini 04PRA}.

In these works 
\cite{Ou97QSemiclOpt,Yurke_Stoler87,Grosshans01,Aichele Lvovsky Schiller 02EurPhysJD,Viciani Zavatta Bellini 04PRA}, 
the parametric down conversion process is treated 
within the first order perturbation theory, 
and 
it is assumed that the idler photon is detected 
by an ideal photon counter. 
The conditional signal state is then measured 
by an imperfect homodyne detector 
with the effective quantum efficiency. 
In the experiment by Wenger et al., however, 
the squeezed state from a parametric amplifier was used, 
which is beyond the first order perturbation theory 
of parametric process. 
The state includes higher photon numbers. 
When the trigger beam from such a state is detected by 
a realistic photon counter 
which can detect the arrival of photons but 
usually fails in precise discrimination of the photon number, 
the conditional output state results in a mixed state. 
Realistic photon counters also suffer from dark counts, 
which cause fake triggers, 
and degrade the quality of the conditional state.

These imperfections were considered in the context 
of observing the negativity of Wigner function distribution 
of a photon-subtracted squeezed state in 
\cite{kim Park Knight Jeong 05PRA}.  
The on/off detector characteristics including the dark counts 
was modeld by the modal purity factor, 
and 
the mode mismatch by the inefficiency of homodyne dtector.  
Unfortunately, however, theories at such a level 
of phenomenological parameterization cannot 
provide practically useful design guidelines. 
In fact, experimentalists want to know, for example, 
an appropriate time duration 
of photon counting and homodyne detection 
in order to observe 
the negativity of Wigner function distribution, 
depending on the spectral characteristics of 
the squeezed state source.

In this paper we develop a multimode theory 
that 
can explicitly calculate the photon-subtracted squeezed state, 
and 
can clarify the conditions 
to observe the negativity of Wigner function distribution 
in realistic situations, 
taking practical imperfections and 
multimode aspects of the quantum states into account. 
Particular emphasis is made on the frequency mode matching 
under the time- and band-limitation to the optical fields.  
Analysis along this line is essential for 
designing the practical setup of the non-Gaussian operations 
and analyzing the measurement results. 
Nevertheless explict analyses have never been given 
so far to our knowledge. 
In fact, this is the most non-trivial aspect in the mode 
matching issue 
\cite{Huang_Kumar89PRA,Zhu_Caves90}. 
This problem is also related with the simultaneous control 
of discrete and continuous natures of quantum optical fields. 
Discrete aspects such as photon counting are often seen 
in the time domain, 
while continuous ones have been exploited most so far 
in the frequency domain. 
The optimal control of both aspects 
relates deeply to the optimal mode matching of 
the time- and band-limited quantum states. 
The necessary theoretical basis was already given by Zhu and Caves 
\cite{Zhu_Caves90}. 
We extend this theory into the scheme of conditional preparation 
of non-Gaussian state 
by photon subtraction from the squeezed state.

In what follows, 
the mode matching in terms of the other degrees of freedom, 
i.e. the spatial and polarization degrees of freedom, 
is neglected for simplicity. 
It is actually made almost perfect in the CW scheme 
based on cavity optical parametric oscillator systems 
by careful cavity lockings. 
We start with a brief review of basic notions 
and mathematical tools on time- and band-limited signals 
in Section 
\ref{Time and band-limited signals}. 
Using this basis, 
we then develop a formulation to describe and analyze 
the mode that the homodyne detector actually observes 
in Section 
\ref{Homodyne detection}. 
In Section 
\ref{On/off detector}, 
we model the on/off detector including practical imperfections. 
In Section 
\ref{Wigner function of the non-Gaussian state}, 
an explicit formulus of the Wigner function distribution 
of the conditional state is given. 
In Section 
\ref{CW scheme}, 
our formalism is applied to the CW scheme, 
and numerical examples of the Wigner function 
and mode matching design charts are given. 
In Section 
\ref{Pulsed scheme}, 
the pulsed scheme is analyzed. 
Section \ref{Conclusion} concludes with a few remarks.

%%%%%%%%%%%%%%%%%%%%%%%%%%%%%%%%%%%%%%%%%%%%%%%%%%%%%%%%%%%%%%%
\section{Time- and band-limited signals}
\label{Time and band-limited signals}
%%%%%%%%%%%%%%%%%%%%%%%%%%%%%%%%%%%%%%%%%%%%%%%%%%%%%%%%%%%%%%%

Consider a light beam 
in a single transverse mode of the optical field 
with a continuous spectrum. 
We denote the positive-frequency part of the field operator 
by 
\beq  
\hat a(t)=\frac{1}{2\pi}\int_{-\infty}^{\infty} d\Omega
\hat a(\omega_0+\Omega)
e^{-i(\omega_0+\Omega)t}. 
\eeq
Here $a(\omega_0+\Omega)$ is the annihilation operator 
for the Fourier component at angular frequency $\omega_0+\Omega$, 
where 
$\omega_0$ is the center angular frequency of 
the spectrum of a laser source. 
The operator obeys the continuum commutation relation 
\beq\label{continuum commutation relation in frequency} 
[\hat a(\omega_0+\Omega), \hat a^\dagger(\omega_0+\Omega')]
=2\pi\delta(\Omega-\Omega'). 
\eeq
The time dependent field operator $\hat a(t)$ is defined in 
the interval $(-\infty,\infty)$, 
and obeys the commutation relation 
\beq\label{continuum commutation relation in time} 
[\hat a(t), \hat a^\dagger(t')]=\delta(t-t'). 
\eeq

Suppose that this transverse mode is excited into 
an ideal band-limited squeezed vacuum state. 
Let the finite bandwidth $B$ in Hz, 
and be centered at the optical frequency $\omega_0/2\pi$. 
We assume that $B\ll\omega_0/2\pi$. 
Such a state is mathematically represented by 
\beq\label{Squeezed vacuum}  
\ket{\mathbf{r}}
=
\hat S 
\ket{\mathbf{0}},  
\eeq
with 
\beqa\label{Squeezing operator}  
\hat S
=\mathrm{exp}
\Biggl(
\frac{1}{2\pi}\int_{-\pi B}^{\pi B} d\Omega
\frac{\gamma(\Omega)}{2}
\biggl[
\hat a(\omega_0+\Omega)\hat a(\omega_0-\Omega)
\nonumber\\
-
\hat a^\dagger(\omega_0+\Omega)\hat a^\dagger(\omega_0-\Omega)
\biggr]
\Biggr), 
\eeqa
where $\gamma(\Omega)=\gamma(-\Omega)$ is 
a frequency-dependent real squeezing parameter. 
This state can be obtained from the output of a degenerate 
parametric amplifier. 
The squeezing bandwidth is determined by the degree of phase 
mismatching. 
A cavity is often used to enhance the nonlinear interaction. 
In such a case the state Eq. (\ref{Squeezed vacuum}) describes 
an ideally simplified output state from the cavity 
provided that the parametric oscillation is operated below 
threshold. 
The bandwidth is then given by the cavity bandwidth. 
Its precise modeling is described in Section 
\ref{CW scheme}.  
By the way, the band limitation becomes essential 
not for describing the squeezing process 
but for analyzing the homodyne current from detector electronics, 
which has a finite, usually not so wide, bandwidth. 
It is this current that defines the measured mode. 
This point is discussed later again.

For mathematical convenience 
we introduce the following operator 
in the rotating frame about 
the center frequency $\omega_0$, 
\beq\label{operator in rotating frame}
\hat A(t)=\hat a(t) e^{i\omega_0t}
=
\frac{1}{2\pi}\int_{-\infty}^{\infty} d\Omega
\hat A(\Omega) e^{-i\Omega t}, 
\eeq
where 
$\hat A(\Omega)=\hat a(\omega_0+\Omega)$. 
The squeezing operator is then rewritten by 
\beqa\label{Squeezing operator in rotating frame}  
\hat S
=\mathrm{exp}
\Biggl(
\frac{1}{2\pi}\int_{-\pi B}^{\pi B} d\Omega
\frac{\gamma(\Omega)}{2}
\biggl[
\hat A(\Omega)\hat A(-\Omega)
\nonumber\\
-
\hat A^\dagger(\Omega)\hat A^\dagger(-\Omega)
\biggr]
\Biggr). 
\eeqa

\begin{figure}
\hspace{10mm}
\begin{center}
\includegraphics[width=1.0\linewidth]{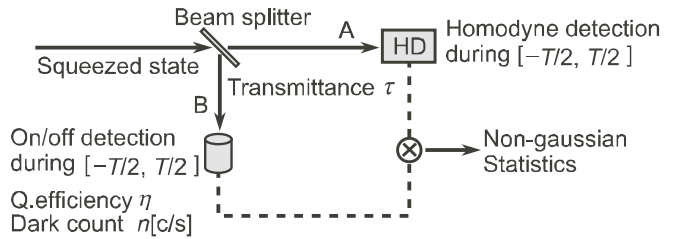}
\caption{Scheme of the non-Gaussian operation induced by 
the measurement by the on/off detector. }
\label{scheme_of_setup}
\end{center}
\end{figure}

Using the squeezed source thus described, 
we now consider a scheme depicted in Fig. \ref{scheme_of_setup}. 
The squeezed vacuum beam is splitted via a beamsplitter 
with a small reflectance $1-\tau$. 
The reflected beam in path B is guided into an on/off detector. 
The main transmitted beam at path A, the signal beam, is 
measured by a homodyne detector. 
The ``on" signals at the on/off detector 
are used as the triggers to select homodyne events. 
It is assumed that 
both the on/off and the homodyne detectors have 
the same measurement time duration $T$. 
It is this time interval $T$ 
to define events in the time domain. 
Actually, 
it is not until 
the measured modes (states) by the on/off and homodyne detectors 
are specified 
that the quantum states of 
the trigger photons and the conditionally selected homodyne events 
are clearly defined.  
In other words, 
it is neither 
the temporal width of the input pulse nor its spectral bandwidth 
that directly determines 
the characteristics of the conditional events.

%%%%%%%%%%%%%%%%%%%%%%%%%%%%%%%%%%%%%%
%%%%%%%%%%%%%%%%%%%%%%%%%%%%%%%%%%%%%%

In order to analyze this scheme 
we should find an appropriate orthonormal basis set 
for describing simultaneously 
the squeezing, 
the homodyne detection process, 
and 
the on/off detection process. 
The squeezing and homodyne detection of it are described 
most easily in the frequency domain, 
whereas the on/off detection is based on 
the photon counting events 
appearing in the time domain. 
The continuum set of the frequency modes $\{\hat A(\Omega)\}$ 
satisfying the commutation relation 
Eq. (\ref{continuum commutation relation in frequency}) 
is not an appropriate set.

Band-limited signals to $[-B/2, B/2]$ in Hz can be 
uniquely represented by 
\beqa\label{expansion by sinc functions}  
\hat A_B(t)
&=&
\frac{1}{2\pi}\int_{-\pi B}^{\pi B} d\Omega
\hat A(\Omega)
\mathrm{e}^{-i\Omega t}, 
\nonumber\\
&=&
\frac{1}{\sqrt{B}}
\sum_{k=-\infty}^{\infty}
\hat A(\frac{k}{B}) \theta_k(t),  
\eeqa
where 
$\theta_k(t)$ is the sinc function 
\beq\label{sinc functions}
\theta_k(t)=\sqrt{B}
\frac{\sin\pi B(t-\frac{k}{B})}{\pi B(t-\frac{k}{B})}, 
\eeq
satisfying the orthonormal relation  
\beq\label{orthonormal relation of sinc functions}
\int_{-\infty}^{\infty}dt 
\theta_k(t) \theta_l(t)
=\delta_{kl}. 
\eeq
This set, however, is not orthogonal on a finite time duration. 
So the use of this set is not convenient 
for describing the events defined on a finite time duration.

Time-limited signals to $[-T/2, T/2]$, 
on the other hand, 
can be uniquely represented by 
\beq\label{expansion by DFT functions}  
\hat A_T(t)
=
\sum_{k=-\infty}^{\infty}
\hat A_{T,k} \phi_k(t),  
\eeq
where 
\beq\label{DFT functions}
\phi_k(t)
=
\left\{
\begin{array}{ll}
\displaystyle\frac{1}{\sqrt{T}} 
\mathrm{exp}(-i\displaystyle\frac{2\pi kt}{T}) 
& \quad 
 -\displaystyle\frac{T}{2}
  \le t \le 
  \displaystyle\frac{T}{2}, 
\\
0 
& \quad 
\mathrm{otherwise},
\end{array}\right. 
\eeq
satisfying the orthonormal relation  
\beq\label{orthonormal relation of DFT functions}
\int_{-T/2}^{T/2}dt 
\phi_k^\ast(t) \phi_l(t)
=\delta_{kl}. 
\eeq
The Fourier coefficient $\hat A_{T,k}$ is defined by 
\beq\label{DFT Fourier coefficient}
\hat A_{T,k}
=\frac{1}{\sqrt{T}} \int_{-T/2}^{T/2}dt 
\hat A(t) \phi_k^\ast(t), 
\eeq
using the operator in 
Eq. (\ref{operator in rotating frame}). 
Although this set is useful to describe the events 
in the time domain, 
it makes the analysis complicated 
when the signal is band-limited. 
Such a band-limited analysis becomes essential 
in describing the homodyne current from detector electronics 
with a finite bandwidth.

The appropriate set to expand the time- and band-limited signals 
is known as prolate spheroidal wave functions 
\cite{Prolate spheroidal wave functions}. 
These functions frequently apprear in diverse problems 
in physics and mathematics, 
and were first introduced to quantum optics by Zhu and Caves 
\cite{Zhu_Caves90}. 
These functions are defined 
by the solutions of the integral equation 
in the time domain  
\beq\label{Integral equation for Psi in text}
\chi_k(c) \Psi_{k}(c,t) 
=\int_{-T/2}^{T/2} dt' 
\frac{\sin \pi B(t-t')}
     {\pi(t-t')}
\Psi_{k}(c,t'),  
\eeq
or equivalently in the frequency domain 
\beq\label{Integral equation for Phi in text}
\chi_k(c) \Phi_{k}(c,\Omega) 
=\int_{-\pi B}^{\pi B} d\Omega' 
\frac{\sin \frac{(\Omega-\Omega')T}{2}}
     {\pi(\Omega-\Omega')}
\Phi_{k}(c,\Omega').  
\eeq
Actually, given any $T>0$ and any $B>0$, 
one can find a countably infinite set of real functions for 
the integral equation 
\beq\label{Integral equation to define S_0k}
\chi_k(c) S_{0k}(c,x) 
=\int_{-1}^1 dy \frac{\sin c(x-y)}{\pi(x-y)}
S_{0k}(c,y), 
\quad
\vert x\vert\le1,  
\eeq 
and a set of real positive numbers satisfying 
\beq
1\ge\chi_0(c)>\chi_1(c)>\chi_2(c)>\cdots,  
\eeq 
where $c=\pi BT/2$. 
The solution $S_{0k}(c,x)$ 
is called the angular prolate spheroidal wave functions 
whose properties are summarized in Appendix 
\ref{Appendix:Prolate Spheroidal Wave Functions}.  
The eigenvalues $\chi_k(c)$ are expressed by using 
a second set of solution $R_{0k}^{(1)}(c,x)$, 
called the radial prolate spheroidal wave functions, 
as 
\beq
\chi_k(c)=\frac{2c}{\pi}\left[R_{0k}^{(1)}(c,1)\right]^2.  
\eeq
$R_{0k}^{(1)}(c,x)$ differ from $S_{0k}(c,x)$ 
only by a real scale factor.

Using the above functions, 
the complete and orthonormal set on the finite bandwidth 
$\vert\Omega\vert\le\pi B$ can be given by 
\beq
\Phi_k(c, \Omega)
=
\sqrt{\frac{2k+1}{B}}S_{0k}(c,\frac{\Omega}{\pi B}), 
\quad
k=0, 1, \cdots. 
\eeq
These functions satisfy 
\beq
\Phi_k(c, \Omega)=(-1)^{k}\Phi_k(c, -\Omega),
\eeq
\beq
\frac{1}{2\pi}
\int_{-\pi B}^{\pi B} d\Omega
\Phi_k(c, \Omega)\Phi_l(c, \Omega)=\delta_{kl}. 
\eeq

The Fourier transforms of the functions $\Phi_k(c, \Omega)$ 
give mode functions in the time domain. 
They are explicitly 
\beqa\label{FT of prolate spheroidal functions}
\Psi_k(c, t)
&=&
\frac{1}{2\pi}\int_{-\pi B}^{\pi B} d\Omega
\Phi_k(c, \Omega) e^{-i\Omega t}
\nonumber\\
&=&\sqrt{(2k+1)B}(-i)^k 
R_{0k}^{(1)}(c,1)
S_{0k}(c,\frac{2t}{T}). 
\eeqa
The functions $\Psi_k(c, t)$ are defined in the interval 
$(-\infty,\infty)$ and are orthonormal, 
\beq\label{Orthonormal condition for Psi}
\int_{-\infty}^{\infty} dt
\Psi_k(c, t)\Psi_l(c, t)^\ast=\delta_{kl}. 
\eeq
They also keep the orthogonality relation 
over the interval $[-T/2, T/2]$, 
\beq\label{Orthogonal condition for Psi}
\int_{-T/2}^{T/2} dt
\Psi_k(c, t)\Psi_l(c, t)^\ast=\chi_k(c)\delta_{kl}. 
\eeq
For a fixed value of $c$ the $\chi_k(c)$ fall off to zero 
rapidly with increasing $k$ once $k$ has exceeded 
$2c/\pi$ $(=BT)$. 
A small value of $\chi_k(c)$ implies that 
$\Psi_k(c, t)$ will have most of its weight 
outside the interval $[-T/2, T/2]$ 
whereas 
a value of $\chi_k(c)$ near 1 implies that 
$\Psi_k(c, t)$ will be concentrated largely in $[-T/2, T/2]$. 
The inverse Fourier transformation is 
\beq\label{FT of prolate spheroidal functions 2 in text}
\chi_k(c)\Phi_k(c, \Omega)
=
\int_{-T/2}^{T/2} dt
\Psi_k(c, t) e^{i\Omega t}.
\eeq

%%%%%%%%%%%%%%%%%%%%%%%%%%%%%%%%%%%%%%
\begin{figure*}
\hspace{10mm}
\begin{center}
\includegraphics[width=0.6\linewidth]{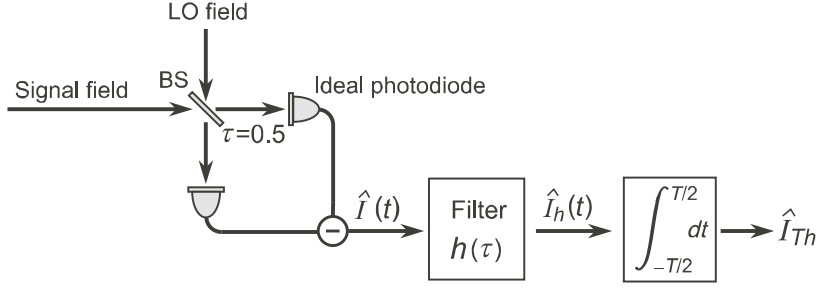}
\caption{Scheme of balanced homodyne detection.}
\label{balanced_homodyne_detector}
\end{center}
\end{figure*}
%%%%%%%%%%%%%%%%%%%%%%%%%%%%%%%%%%%%%%

The time- and band-limited fields are now quantized 
in terms of these mode $k$. 
We thus introduce the discrete set of the operators 
\beqa\label{definition of A_k}  
\hat A_k 
&=&
\frac{1}{2\pi}\int_{-\pi B}^{\pi B} d\Omega
\hat A(\Omega) \Phi_k(c, \Omega), 
\nonumber\\
&=&
\int_{-\infty}^{\infty}dt
\hat A(t) \Psi_k^\ast(c, t),  
\eeqa
which obeys the commutation relation 
\beq\label{commutation relation of A_k} 
[\hat A_k, \hat A_l^\dagger]
=\delta_{kl}. 
\eeq
The operators $\hat A(\Omega)$ confined in the finite bandwidth $B$, 
such as the ones appearing in the squeezing operator 
Eq. (\ref{Squeezing operator in rotating frame}), 
can be expanded as 
\beq\label{A(Omega) expansion by A_k}
\hat A(\Omega)=\sum_{k=0}^{\infty} \hat A_k \Phi_k(c, \Omega).  
\eeq
Let us assume a rectangular squeezing spectrum 
$\gamma(\Omega)=\gamma$ 
in Eq.(\ref{Squeezing operator in rotating frame}). 
The squeezing operator then factors into 
a product of single mode squeezing operators 
\beq\label{Flat band squeezing operator}  
\hat S
=\bigotimes_{k=0}^{\infty}
\hat S_k, 
\quad 
\hat S_k=
\mathrm{exp}
\left[
  \frac{r_k}{2}
  \left( \hat A_k^2 - \hat A_k^{\dagger2} \right)
\right], 
\eeq
with the squeezing parameter for mode $k$ 
\beq\label{r_k}  
r_k=(-1)^k\gamma. 
\eeq
The multimode squeezed vacuum state Eq. (\ref{Squeezed vacuum}) 
is now represented by 
\beq\label{Squeezed vacuum2}
\ket{\mathbf{r}}
=\bigotimes_{k=0}^{\infty}\ket{r_k}, 
\quad 
\ket{r_k}=\hat S_k \ket{0}. 
\eeq

%%%%%%%%%%%%%%%%%%%%%%%%%%%%%%%%%%%%%%%%%%%%%%%%%%%%%%%%%%%%%%%
\section{Homodyne detection}
\label{Homodyne detection}
%%%%%%%%%%%%%%%%%%%%%%%%%%%%%%%%%%%%%%%%%%%%%%%%%%%%%%%%%%%%%%%

In order to analyze the mode matching issue in the non-Gaussian 
operation depicted in Fig. \ref{scheme_of_setup}, 
it is essential to know what kind of modes are actually 
measured in the detectors. 
In this section we study the homodyne detection 
from this point of view. 
The most commonly used homodyne detection scheme 
is the so-called balanced homodyne detector. 
Its typical scheme is shown in Fig. 
\ref{balanced_homodyne_detector}. 
In this scheme 
the signal field $\hat A(t)$ is first combined 
with the LO field $\hat A^\mathrm{L}(t)$, 
via a 50:50 beam splitter 
and the two output beams are converted into photocurrents 
at each photodetector. 
The photocurrents from the two detectors are balanced 
to produce the difference current.

If the photodetector has a $\delta$-function response in time 
so that it covers the infinite bandwidth, 
the photocurrents are produced intantaneously. 
The difference current can then be given by 
\beq\label{Difference homodyne current}
\hat I(t)
=
\hat A^{\mathrm{L}\dagger}(t) \hat A(t)
+\hat A^\dagger(t) \hat A^\mathrm{L}(t). 
\eeq
In practical case, however, 
photodetectors themselves have a non-$\delta$-function response 
and have a finite bandwidth. 
Furthermore the difference current is often electrically 
amplified, and is finally analyzed. 
The bandwidth of homodyne detector electronics is also 
limited, usually to a few hundred MHz at most. 
These effects can be modeled by the response function $h(t)$ 
such that the instantaneous difference current is filtered 
through this function 
\cite{Yurke85,Ou_Kimble95}. 
The operator for the filtered current is given by 
\beq\label{filtered homodyne current}
\hat I_h(t)
=
\int_{-\infty}^\infty d\tau h(t-\tau) \hat I(\tau). 
\eeq
This current is integrated over the time duration 
$[-T/2, T/2]$ 
\beq\label{integrated homodyne current after filtering}
\hat I_{Th}
=
\int_{-T/2}^{T/2} dt \hat I_h(t). 
\eeq
This operator directly corresponds to the final observable 
in the homodyne detection in the present context. 
Explicit calculation of this current $\hat I_{Th}$ will be made 
for each specific physical model in later sections.

In this section we first assume 
an ideal homodyne detector with the infinite bandwidth, 
i.e. $h(t-\tau)=\delta(t-\tau)$, 
and derive an explicit expression for the final observable. 
Since the LO field is usually a strong classical field 
so that its quantum fluctuations can be neglected. 
We further assume that the LO field has 
the same or narrower bandwidth than that of the squeezing 
so that the LO field can be band-limited at least to 
$\vert\Omega\vert\le\pi B$. 
One may then expand it as 
\beq
\hat A^\mathrm{L}(t)
=\sum_{k=0}^{\infty} \hat A_k^\mathrm{L} \Psi_k(c, t)
=\sum_{k=0}^{\infty} \alpha_k^\mathrm{L} 
e^{i\phi}\Psi_k(c, t).  
\eeq

One should, on the other hand, note that 
the field operator at the signal port $\hat A(t)$ in 
Eq. (\ref{Difference homodyne current}) 
can NOT be band-limited 
because the ideal homodyne detector 
can be sensitive to the modes even outside the bandwidth 
of the input squeezed beam, 
and these modes add the vacuum fluctuations 
to the homodyne current. 
So we decompose the field $\hat A(t)$ into 
a sum of the following two components  
\beq
\hat A(t)=\hat A_B(t)+\hat A_V(t).  
\eeq
Here the first component $\hat A_B(t)$ is band-limited to 
the squeezing bandwidth, 
and hence can be expanded as 
\beq\label{A_B(t)}
\hat A_B(t)
=
\frac{1}{2\pi}\int_{-\pi B}^{\pi B} d\Omega
\hat A(\Omega) e^{-i\Omega t}
=
\sum_{k=0}^{\infty} \hat A_k \Psi_k(c, t).   
\eeq
The second component $\hat A_V(t)$ 
consists of the remaining frequency modes 
\beq
\hat A_V(t)
=\frac{1}{2\pi}
\left[
   \int_{-\infty}^{-\pi B} d\Omega
 + \int_{\pi B}^{\infty} d\Omega   
\right]
\hat A(\Omega) e^{-i\Omega t}. 
\eeq
%
%
%
%\beqa
%\hat A_V(t)
%&\equiv&
%\frac{1}{2\pi}\int_V d\Omega
%\hat A(\Omega) e^{-i\Omega t}
%\nonumber\\
%&\equiv&
%\frac{1}{2\pi}
%\left[
%   \int_{-\infty}^{-\pi B} d\Omega
% + \int_{\pi B}^{\infty} d\Omega   
%\right]
%\hat A(\Omega) e^{-i\Omega t}. 
%\eeqa
%
%
%
Accordingly the whole Hibert space is divided into 
two subspaces: 
one is $\cal H^\mathrm{S}$  for the field modes 
inside the squeezing bandwidth $\vert\Omega\vert\le\pi B$, 
and the other $\cal H^\mathrm{V}$ for the field modes 
outside the squeezing bandwidth $\vert\Omega\vert>\pi B$. 
The operators $\hat A_B(t)$ and $\{ \hat A_k \}$ act on 
$\cal H^\mathrm{S}$, 
while 
the operator $\hat A_V(t)$ acts on $\cal H^\mathrm{V}$.

The ideal homodyne current is then written as 
\beqa
\hat I(t)
&=&
\sum_{k=0}^\infty \sum_{l=0}^\infty
\alpha_k^\mathrm{L} e^{-i\phi}
\Psi_k^\ast(c, t)
\hat A_l 
\Psi_l(c, t)
\nonumber\\
&+&
\sum_{k=0}^\infty
\alpha_k^\mathrm{L} e^{-i\phi} 
\Psi_k^\ast(c, t) \hat A_V(t) 
+\mathrm{h.\, c.}, 
\eeqa
which is defined on the whole Hilbert space 
$\cal H^\mathrm{S}\otimes\cal H^\mathrm{V}$. 
This current is integrated over the time duration $T$ 
to define the event signals in the time domain 
\beqa\label{integrated homodyne current by A and A dagger}
\hat I_T
&=&
\int_{-T/2}^{T/2} dt \hat I(t)
\nonumber\\
&=&
\sqrt{2} \sum_{k=0}^\infty
\alpha_k^\mathrm{L} 
\chi_k(c) 
\hat A_k e^{-i\phi}
\nonumber\\
&+&
\sqrt{2} \sum_{k=0}^\infty
\alpha_k^\mathrm{L} 
\sqrt{\chi_k(c) [1-\chi_k(c)]}
\hat V_k e^{-i\phi}
\nonumber\\
&+&
\mathrm{h.\, c.}, 
\eeqa
where we have used the orthogonality relation 
Eq. (\ref{Orthogonal condition for Psi}), 
and 
have introduced the quadrature operators on $\cal H^\mathrm{V}$ 
\beq\label{V_k}
\hat V_{k}
=
\frac{1}{\sqrt{\chi_k(c) [1-\chi_k(c)]}}
\int_{-T/2}^{T/2} dt 
\hat A_V(t) \Psi_k^\ast(c, t),  
\eeq
which satisfies the commutation relation 
\beq\label{commutation relation of V_k} 
[\hat V_k, \hat V_l^\dagger]
=\delta_{kl}. 
\eeq
Define the quadrature operators on $\cal H^\mathrm{S}$ 
for the modes inside the squeezing bandwidth 
$\vert\Omega\vert\le\pi B$ 
\beq\label{quadrature for squeezed mode k}
\hat X_k^\mathrm{S}(\phi)
=
\frac{1}{\sqrt2}
(\hat A_{k} e^{-i\phi} + \hat A_{k}^{\dagger} e^{i\phi}), 
\eeq
and the ones on $\cal H^\mathrm{V}$ 
for the modes outside the squeezing bandwidth 
$\vert\Omega\vert>\pi B$
\beq\label{quadrature for vacuum mode k}
\hat X_k^\mathrm{V}(\phi)
=
\frac{1}{\sqrt2}
(\hat V_{k} e^{-i\phi} + \hat V_{k}^{\dagger} e^{i\phi}), 
\eeq
and 
rewrite 
Eq. (\ref{integrated homodyne current by A and A dagger}) 
as 
\beqa\label{integrated homodyne current by quadrature}
\hat I_T
&=&
\int_{-T/2}^{T/2} dt \hat I(t)
\nonumber\\
&=&
\sqrt{2} \sum_{k=0}^\infty
\alpha_k^\mathrm{L} 
\chi_k(c) 
\hat X_k^\mathrm{S}(\phi)
\nonumber\\
&+&
\sqrt{2} \sum_{k=0}^\infty
\alpha_k^\mathrm{L} 
\sqrt{\chi_k(c) [1-\chi_k(c)]}
\hat X_k^\mathrm{V}(\phi). 
\eeqa
This integrated homodyne current operator 
defines the mode which is actually observed 
in the time domain analysis, 
the so called LO matched mode 
\cite{Smithey93_PRL_DC_Homodyne,Raymer95_JOSAB_HomodyneModeMatch}.

We define the quadrature operator 
for the LO matched mode in the following way. 
Define the weight factors 
\beq\label{w_k^S} 
w_k^\mathrm{S}
=
\frac
{\alpha_k^\mathrm{L} \chi_k(c)}
{\sqrt{\displaystyle\sum_{k=0}^\infty 
       (\alpha_k^\mathrm{L})^2 \chi_k(c)}}, 
\eeq
and 
\beq\label{w_k^V} 
w_k^\mathrm{V}
=
\frac
{\alpha_k^\mathrm{L} {\sqrt{\chi_k(c) [1-\chi_k(c)]}}}
{\sqrt{\displaystyle\sum_{k=0}^\infty 
       \left(\alpha_k^\mathrm{L}\right)^2 \chi_k(c)}},  
\eeq
such that 
\beq\label{normalization of w_k} 
\sum_{k=0}^\infty 
\left[
\left( w_k^\mathrm{S} \right)^2
+
\left( w_k^\mathrm{V} \right)^2
\right]
=1.  
\eeq
In order to simplify the notation on the Hilbert space 
$\cal H^\mathrm{S}\otimes\cal H^\mathrm{V}$,  
let us further introduce vector notations 
\beq\label{vector of quadrature X_phi}
\hat{\mathbf{X}}(\phi)
\equiv
\left[
  \begin{array}{c}
  \hat X_0(\phi) \\
  \hat X_1(\phi) \\
  \hat X_2(\phi) \\
  \vdots     \\
  \end{array}
\right]
\equiv
\left[
  \begin{array}{c}
  \hat X_0^\mathrm{S}(\phi) \\
  \hat X_1^\mathrm{S}(\phi) \\
  \vdots     \\
  \hat X_0^\mathrm{V}(\phi) \\
  \hat X_1^\mathrm{V}(\phi) \\
  \vdots     \\
  \end{array}
\right],
\eeq
\beq\label{vector epsilon}
\bm{\epsilon}
\equiv
\left[
  \begin{array}{c}
  \epsilon_0 \\
  \epsilon_1 \\
  \epsilon_2 \\
  \vdots     \\
  \end{array}
\right]
\equiv
\left[
  \begin{array}{c}
  w_0^\mathrm{S} \\
  w_1^\mathrm{S} \\
  \vdots     \\
  w_0^\mathrm{V} \\
  w_1^\mathrm{V} \\
  \vdots     \\
  \end{array}
\right], 
\eeq
and 
\beq\label{vector of eigenvalue x_phi}
{\mathbf{x}}(\phi)
\equiv
\left[
  \begin{array}{c}
  x_0(\phi) \\
  x_1(\phi) \\
  x_2(\phi) \\
  \vdots     \\
  \end{array}
\right]
\equiv
\left[
  \begin{array}{c}
  x_0^\mathrm{S}(\phi) \\
  x_1^\mathrm{S}(\phi) \\
  \vdots     \\
  x_0^\mathrm{V}(\phi) \\
  x_1^\mathrm{V}(\phi) \\
  \vdots     \\
  \end{array}
\right],
\eeq
where 
\beq\label{eigenequation for mode k on H^S}
\hat X_k^\mathrm{S}(\phi) \ket{x_k^\mathrm{S}(\phi)}
=x_k^\mathrm{S}(\phi) \ket{x_k^\mathrm{S}(\phi)}, 
\eeq
and 
\beq\label{eigenequation for mode k on H^V}
\hat X_k^\mathrm{V}(\phi) \ket{x_k^\mathrm{V}(\phi)}
=x_k^\mathrm{V}(\phi) \ket{x_k^\mathrm{V}(\phi)}. 
\eeq
Here we have 
\beq
[\hat X_k(\phi), \hat X_l(\phi+{\pi\over2})]=i\delta_{kl}. 
\eeq
The integrated homodyne current 
Eq. (\ref{integrated homodyne current by quadrature}) 
is then written as 
\beq\label{integrated homodyne current by LO matched quadrature}
\hat I_T
=
{\sqrt{2\displaystyle\sum_{k=0}^\infty 
\left(\alpha_k^\mathrm{L}\right)^2 \chi_k(c)}}
\tilde X_0(\phi)
\eeq
where 
\beq\label{observed quadrature}
\tilde X_0(\phi)
=
{}^t \bm{\epsilon} \cdot \hat{\mathbf{X}}(\phi)
\eeq
is the quadrature operator for the LO matched mode, 
which is defined on the entire Hilbert space 
$\cal H^\mathrm{S}\otimes\cal H^\mathrm{V}$. 
It is this operator 
that corresponds to the final observable 
in the homodyme detector 
for the time domain non-Gaussian operation. 
In the following calculation 
we only need quadrature operators at $\phi=0$, 
those are simply represented by 
\beqa
\hat X_k&=&\hat X_k(0),
\nonumber\\
\hat P_k&=&\hat X_k({\pi\over2}),
\nonumber\\
\tilde X_0&=&\tilde X_0(0),
\nonumber\\
\tilde P_0&=&\tilde X_0({\pi\over2}).
\nonumber\\ 
\eeqa

The POVM and the resulting measurement statistics 
for the integrated homodyne current is described 
by using the eigenstates of $\tilde X_0$ 
\beq\label{tilde eigenstate of mode 0}
\tilde X_0\varket{x_0'}=x_0'\varket{x_0'}.  
\eeq
The round ket notation $\varket{\,\,\,}$ is used 
to discriminate these eigenstates 
from the ones for the quadrature operators 
which describe the input squeezed 
$\{ \hat X_0^\mathrm{S}, \hat X_1^\mathrm{S}, \cdots \}$ 
and 
the vacuum 
$\{ \hat X_0^\mathrm{V}, \hat X_1^\mathrm{V}, \cdots \}$ 
fields 
\beq\label{hat eigenstate of mode k}
\hat X_k\ket{x_k}=x_k\ket{x_k}, \quad(k=0,1,\cdots).     
\eeq

In order to represent the quantum state of measured mode, 
and to calculate the corresponding Wigner function, 
we have to derive the formula to connect 
$\{\varket{x_0'}\}$ with $\{\ket{x_0}, \ket{x_1}, \cdots \}$. 
For this purpose we consider a real unitary transformation 
\beq\label{unitary transformation matrix} 
\left[
  \begin{array}{c}
  \tilde X_0 \\ \tilde X_1 \\ \tilde X_2 \\ \vdots 
  \end{array}
\right]
=
\left[
  \begin{array}{cccc}
  \epsilon_0 & \epsilon_1 & \epsilon_2 & \cdots \\
  u_{10}     & u_{11}     & u_{12}     & \cdots \\
  u_{20}     & u_{21}     & u_{22}     & \cdots \\
  \vdots     & \vdots     & \vdots     & \ddots \\
  \end{array}
\right]
\left[
  \begin{array}{c}
  \hat X_0 \\ \hat X_1\\ \hat X_2 \\ \vdots 
  \end{array}
\right],  
\eeq
which includes the linear relation of 
Eq. (\ref{observed quadrature}) 
in the 0th component.  
We denote this equation as 
\beq
\tilde{\mathbf{X}}= {\mathbf{U}} \hat{\mathbf{X}}.  
\eeq
Such a unitary transformation is not unique, 
but here we need not to know its explicit components. 
As in Eq. (\ref{tilde eigenstate of mode 0}), 
the eigenstates of $\tilde X_k$ are denoted by 
\beq\label{tilde eigenstate of mode k}
\tilde X_k\varket{x_k'}=x_k'\varket{x_k'}, \quad(k=0,1,\cdots). 
\eeq
The two sets of operators $\{\hat X_k\}$ and $\{\tilde X_k\}$ 
can be considered to act on the same Hilbert space 
$\cal H^\mathrm{S}\otimes\cal H^\mathrm{V}$. 
In fact, 
the tensor product states of all the modes 
for the two sets of eigenstates 
%$\ket{\mathbf{x}}$ and $\varket{\mathbf{x'}}$ 
are related with each other by 
\beqa\label{ket transformation}
\ket{\mathbf{x}}
&\equiv&
\bigotimes_{k=0}^\infty \ket{x_k}
\nonumber\\
&=&\exp(-i{}^t\mathbf{x}\cdot\hat\mathbf{P})\ket{\mathbf{0}}
\nonumber\\
&=&\exp(-i{}^t\mathbf{x} \mathbf{U}^\dagger
          \cdot
          \mathbf{U} \hat\mathbf{P})\ket{\mathbf{0}}
\nonumber\\
&=&\exp(-i{}^t\mathbf{x'}\cdot\tilde\mathbf{P})\ket{\mathbf{0}}
\nonumber\\
&=&
\bigotimes_{k=0}^\infty \varket{x'_k}
\nonumber\\
&=&\varket{\mathbf{x'}},  
\eeqa
where 
\beq\label{x'=Ux}
\mathbf{x'}=\mathbf{U}\mathbf{x}. 
\eeq

Given an input multimode field in a quantum state $\hat\rho$ 
on $\cal H^\mathrm{S}\otimes\cal H^\mathrm{V}$, 
the integrated homodyne current delivers information 
only on the LO matched mode in $\hat\rho$. 
Therefore what we actually observe is a reduced density 
operator on the subspace $\cal H^\mathrm{L}$ 
spanned by $\varket{x_0'}$ 
\beq\label{reduced density operator}
\tilde\rho
\equiv
\int dx'_1 \int dx'_2 \cdots 
\varbra{x'_1, x'_2,  \cdots}
\hat\rho
\varket{x'_1, x'_2,  \cdots}. 
\eeq 
The matrix element in terms of the quadrature eigenstates 
for the LO matched mode on the subspace $\cal H^\mathrm{L}$ 
is given by 
\beqa\label{matrix element1}
\varbra{x}\tilde\rho\varket{y}
&=&
\int dx'_0 \int dx'_1 \int dx'_2 \cdots 
\delta(x-x'_0)
\nonumber\\
&{}&\times
\varbra{x'_0, x'_1, x'_2,  \cdots}
\hat\rho
\varket{y, x'_1, x'_2,  \cdots}
\nonumber\\
&=&
\int d\mathbf{x'} 
\delta(x-{}^t\bm{\epsilon} \cdot \mathbf{x})
\nonumber\\
&{}&\times
\varbra{\mathbf{x}'}
\hat\rho
\varket{y, x'_1, x'_2,  \cdots}, 
\eeqa 
where we have used the relation of the eigenvalues 
for the LO matched mode 
\beq
x'_0=\sum_{l=0}^\infty \epsilon_l x_l
    ={}^t \bm{\epsilon} \cdot \mathbf{x}. 
\eeq
We then transform the variables 
$\mathbf{x'}={}^t(x'_0, x'_1, x'_2,  \cdots)$ 
into $\mathbf{x}={}^t(x_0, x_1, x_2,  \cdots)$ 
by Eq. (\ref{x'=Ux}).  
Firstly we have 
\beq
\int d\mathbf{x'}=\int d\mathbf{x}, 
\eeq
because the Jacobian of this variable transformation is 
$\vert\mathrm{det}\mathbf{U}\vert=1$. 
Secondly 
\beq
\varbra{\mathbf{x}'}
=\bra{\mathbf{x}}
\eeq
according to Eq. (\ref{ket transformation}). 
Thirdly the quadrature eigenstates 
$\varket{y, x'_1, x'_2,  \cdots}$ 
is converted as 
\beqa
\varket{y, x'_1, x'_2,  \cdots}
&=&
\bigotimes_{k=0}^\infty 
\ket{x_k + 
\bigl( 
      y-{}^t \bm{\epsilon} \cdot \mathbf{x} 
\bigr) 
\epsilon_k} 
\nonumber\\
&=&\ket{\mathbf{x}+
(y-{}^t \bm{\epsilon} \cdot \mathbf{x})
\bm{\epsilon}}, 
\eeqa
because  
\beq\label{unitary transformation of eigenvalues 1} 
\left[
  \begin{array}{c}
  y \\ x_1' \\ x_2' \\ \vdots 
  \end{array}
\right]
=
\mathbf{U}
\left[
  \begin{array}{c}
  x_0 + (y-x_0') \epsilon_0 \\ 
  x_1 + (y-x_0') \epsilon_1 \\ 
  x_2 + (y-x_0') \epsilon_2 \\ 
  \vdots 
  \end{array}
\right],  
\eeq
which can be obtained by eliminating the terms 
$\sum_{k=1}^\infty u_{kj} x_k'$ 
in the following two equations 
\beq\label{unitary transformation of eigenvalues 2} 
\mathbf{U}^\dagger
\left[
  \begin{array}{c}
  y \\ x_1' \\ x_2' \\ \vdots 
  \end{array}
\right]
=
\left[
  \begin{array}{c}
  \epsilon_0 y + \sum_{k=1}^\infty u_{k0} x_k' \\ 
  \epsilon_1 y + \sum_{k=1}^\infty u_{k1} x_k' \\ 
  \epsilon_2 y + \sum_{k=1}^\infty u_{k2} x_k' \\ 
  \vdots 
  \end{array}
\right],  
\eeq
and 
\beq\label{unitary transformation of eigenvalues 3} 
\left[
  \begin{array}{c}
  x_0 \\ x_1 \\ x_2 \\ \vdots 
  \end{array}
\right]
=
\mathbf{U}^\dagger
\left[
  \begin{array}{c}
  x_0' \\ x_1' \\ x_2' \\ \vdots 
  \end{array}
\right]
=
\left[
  \begin{array}{c}
  \epsilon_0 x_0' + \sum_{k=1}^\infty u_{k0} x_k' \\ 
  \epsilon_1 x_0' + \sum_{k=1}^\infty u_{k1} x_k' \\ 
  \epsilon_2 x_0' + \sum_{k=1}^\infty u_{k2} x_k' \\ 
  \vdots 
  \end{array}
\right],  
\eeq
The matrix element 
Eq. (\ref{matrix element1}) is finally given by 
%%%%%%%%%%%%%%%%%%%%%%%%%%%%%%%%%%%%%%%%%%%%%%%%%%%%%%%%%%%%
\beq\label{matrix element2}
\varbra{x}\tilde\rho\varket{y}
=\int d\mathbf{x} 
\delta(x-{}^t \bm{\epsilon} \cdot \mathbf{x})
\bra{\mathbf{x}}
\hat\rho
\ket{\mathbf{x}+(y-x)\bm{\epsilon}}.
\eeq
%%%%%%%%%%%%%%%%%%%%%%%%%%%%%%%%%%%%%%%%%%%%%%%%%%%%%%%%%%%%
This formulus allows one to calculate the statistics of 
the LO matched mode in the time-integrated homodyne detection 
(the left-hand side), 
using the quantum state originally represented 
in terms of the plorate spheroidal wave function modes 
(the right-hand side).

%%%%%%%%%%%%%%%%%%%%%%%%%%%%%%%%%%%%%%%%%%%%%%%%%%%%%%%%%%%%%%%
\section{On/off detector}
\label{On/off detector}
%%%%%%%%%%%%%%%%%%%%%%%%%%%%%%%%%%%%%%%%%%%%%%%%%%%%%%%%%%%%%%%

In contrast to that homodyne detectors can be implemented 
in the near ideal condition at least for the near infrared 
wavelengths at present using Si p-i-n photodiodes, 
on/off detectors usually suffer from imperfect efficiency 
and dark counts. 
In the present context 
where an on/off detector is used to select events, 
dark counts essentially influence the quality of 
the non-Gaussian operation.

The number operator for photons arriving at the detector 
during the interval $T$ is 
\beqa
\hat n_T
&=&
\int_{-T/2}^{T/2} dt \hat a^\dagger(t) \hat a(t) 
\nonumber\\
&=&
\int_{-T/2}^{T/2} dt \hat A^\dagger(t) \hat A(t). 
\eeqa 
Although the signal field $\hat A(t)$ 
appearing in the homodyne current cannot be band-limited 
to the squeezing bandwidth,  
it CAN be here, 
because the vacuum field components outside the squeezing bandwidth 
does not induce photon counts. 
So $\hat A(t)$ can be replaced by $\hat A_B(t)$, 
and can be expanded as in Eq. (\ref{A_B(t)}). 
We then obtain 
\beq
\hat n_T
=
\sum_{k=0}^\infty \chi_k(c) \hat A_k^\dagger\hat A_k. 
\eeq

For mode $k$ we define 
\beq
\hat n_k=\hat A_k^\dagger\hat A_k, 
\quad
\hat n_k \ket{n_k} = n_k \ket{n_k}. 
\eeq
For a photon counter with a finite quantum efficiency $\eta_k$ 
for mode $k$, 
the POVM element registering $n$ photocarriers due to 
photons in mode $k$ is given by 
\cite{Barnett98_PhptonCounter},  
\beq
\hat{\Pi}_k (n;\eta_k)=\sum _{m=n}^\infty 
\left( \begin{array}{c} m \cr n \end{array} \right)
\eta_k^n (1-\eta_k)^{m-n} |m\rangle \langle m|. 
\eeq
The effective quantum efficiency $\eta_k$ for mode $k$ here 
is of the form $\eta_k=\eta\chi_k(c)$, 
where 
$\eta$ is the net detection efficiency 
determined by the total coupling efficiency of photons through 
optical components to the photondetector 
and the intrinsic quantum efficiency of the photondetector. 
This is assumed to be constant over the squeezing bandwidth. 
The factor 
$\chi_k(c)$ represents the weight of mode $k$ on the counting 
interval $[-T/2, T/2]$. 
Taking the effect of the dark counts into account, 
the POVM registering $n$ counts is then given by 
\beq
\hat{\Pi}_k(n;\eta_k,\nu_k)=
\sum_{n'=0}^n 
e^{-\nu_k} 
\frac{\nu_k^{n-n'} }{(n-n')!} \hat{\Pi}_k (n';\eta_k), 
\eeq
where $\nu_k$ is the mean number of the dark counts 
for photons in mode $k$. 
(The dark counts may occur regardless of the bandwidth 
of the input signal field, however, 
they can be taken into account by adjusting the values $\nu_k$ 
for the band-limited modes 
such that the total dark count rate in practical detecotrs 
is properly modeled.) 
These elements satisfy the completeness relation for each mode $k$, 
\begin{equation}
\sum _{n=0} ^\infty \hat{\Pi}_k(n;\eta_k,\nu_k)=
\hat{I}_k. 
\label{completeness}
\end{equation}

The multimode on/off detector placed at path B 
in Fig. \ref{scheme_of_setup} is finally modeled by a binary POVM 
with parameters 
$\bm{\eta}=(\eta_0,\eta_1,\cdots)$ 
and 
$\bm{\nu}=(\nu_0,\nu_1,\cdots)$, 
\beqa\label{on/off POVM}
\hat{\Pi}_\mathrm{off}^\mathrm{B}(\bm{\eta},\bm{\nu})
&=&\bigotimes_{k=0}^\infty 
\hat{\Pi}_k^\mathrm{B} (0;\eta_k,\nu_k), 
\nonumber\\
\hat{\Pi}_\mathrm{on}^\mathrm{B}(\bm{\eta},\bm{\nu})
&=&
\hat I^\mathrm{B}
-\hat{\Pi}_\mathrm{off}^\mathrm{B}(\bm{\eta},\bm{\nu}). 
\eeqa

%%%%%%%%%%%%%%%%%%%%%%%%%%%%%%%%%%%%%%%%%%%%%%%%%%%%%%%%%%%%%%%
\section{Wigner function of the non-Gaussian state}
\label{Wigner function of the non-Gaussian state}
%%%%%%%%%%%%%%%%%%%%%%%%%%%%%%%%%%%%%%%%%%%%%%%%%%%%%%%%%%%%%%%

We now explicitly calculate the non-Gaussian output state 
conditionally obtained by the on/off detector 
specified in the previous section, 
and derive an expression of the Wigner function 
measured by the ideal homodyne detector described in Section 
\ref{Homodyne detection}. 
For convenience of mathematical handling, 
we express the squeezed vacuum state 
\beq\label{Squeezed vacuum3}
\ket{\mathbf{r}^\mathrm{S}}
=\bigotimes_{k=0}^{\infty}\ket{r_k}, 
\quad 
\ket{r_k}=\hat S_k \ket{0}, 
\eeq
in terms of the coherent state basis 
$\ket{\alpha_k}$ 
on the Hibert space $\cal H^\mathrm{S}$. 
The component of mode $k$ can be expanded as  
\beqa\label{squeezed vacuum expanded by coherent states}
\ket{r_k}
&=&
%\frac{\sqrt[4]{1-{\lambda_k}^2}}{\pi}
\frac{(1-{\lambda_k}^2)^{1\over4}}{\pi}
\nonumber\\
&\times&
\int_{-\infty}^{\infty}d^2\alpha_k
\exp\left[-\frac{|\alpha_k|^2}{2}
          -\frac{\lambda_k}{2}\alpha_k^{\ast 2}
    \right]
\ket{\alpha_k},
\eeqa
with $\lambda_k=\tanh r_k$. 
As in Fig. \ref{scheme_of_setup}, 
the squeezed vacuum state is beamsplitted into path A and B, 
resulting in a state 
\beqa\label{beam splitted state}
\ket{\rho^\mathrm{S}}_\mathrm{AB}
=
\bigotimes_{k=0}^{\infty}
%\frac{\sqrt[4]{1-{\lambda_k}^2}}{\pi}
\frac{(1-{\lambda_k}^2)^{1\over4}}{\pi}
\int_{-\infty}^{\infty}d^2\alpha_k
\exp\biggr[-\frac{|\alpha_k|^2}{2}
\nonumber\\
          -\frac{\lambda_k}{2}\alpha_k^{\ast 2}
    \biggr]
\ket{\sqrt{\tau}\alpha_k}_\mathrm{A}
\ket{-\sqrt{1-\tau}\alpha_k}_\mathrm{B}, 
\eeqa
where $\tau$ is the transmittance.

The photon subtracted non-Gaussian state at path A is given by 
\beq
\hat\rho_\mathrm{A}^\mathrm{S}
=
\frac{1}{P_\mathrm{det}}
\tr_\mathrm{B}
\left[ 
\ket{\rho^\mathrm{S}}_\mathrm{(AB)}\bra{\rho^\mathrm{S}}
\hat I^\mathrm{A}\otimes\hat{\Pi}_\mathrm{on}^\mathrm{B}
(\bm{\eta},\bm{\nu})
\right],  
\eeq
where 
\beq
P_\mathrm{det}
=
\tr_\mathrm{AB}
\left[ 
\ket{\rho^\mathrm{S}}_\mathrm{(AB)}\bra{\rho^\mathrm{S}}
\hat I^\mathrm{A}\otimes\hat{\Pi}_\mathrm{on}^\mathrm{B}
(\bm{\eta},\bm{\nu})
\right],  
\eeq
is the probability of having the ``on" signals. 
Using Eqs. (\ref{beam splitted state}) and (\ref{on/off POVM}), 
$\hat\rho_\mathrm{A}^\mathrm{S}$ can be represented as 
%\beq
%\hat\rho_\mathrm{A}
%=
%\frac{1}{P_\mathrm{det}}
%\left( 
%\bigotimes_{k=0}^\infty \hat Q_k 
%-
%\bigotimes_{k=0}^\infty \hat R_k 
%\right),  
%\eeq
\beq\label{rho_A}
\hat\rho_\mathrm{A}^\mathrm{S}
=
\frac{1}{P_\mathrm{det}}
\left( 
\bigotimes_{k=0}^\infty \hat R_k(0,0) 
-
\bigotimes_{k=0}^\infty \hat R_k(\eta_k,\nu_k) 
\right),  
\eeq
with 
%\beqa\label{Q_k}
%\hat Q_k
%&=&
%\frac{\sqrt{1-{\lambda_k}^2}}{\pi^2} 
%\int d^2\alpha_k \int d^2\beta_k 
%\exp\Biggl[
%-\frac{2-\tau}{2}
%\nonumber\\
%&\times&
%\biggl( |\alpha_k|^2+|\beta_k|^2 \biggr)
%+(1-\tau)\beta_k^\ast \alpha_k
%\nonumber\\
%&-&\frac{\lambda_k}{2}(\alpha_k^{\ast2}-\beta_k^{2})
%\Biggr] 
%\ket{{\sqrt\tau}\alpha_k}_\mathrm{A}\bra{{\sqrt\tau}\beta_k},
%\eeqa
%and 
\beqa\label{R_k}
\hat R_k(\eta_k,\nu_k)
&=&
\frac{\sqrt{1-{\lambda_k}^2}}{\pi^2} 
\int d^2\alpha_k \int d^2\beta_k 
\exp\Biggl[
-\frac{2-\tau}{2}
\nonumber\\
&\times&
\biggl( |\alpha_k|^2+|\beta_k|^2 \biggr)
+(1-\tau)(1-\eta_k)\beta_k^\ast \alpha_k
\nonumber\\
&-&\frac{\lambda_k}{2}(\alpha_k^{\ast2}+\beta_k^{2})
\nonumber\\
&-&\nu_k
\Biggr] 
\ket{{\sqrt\tau}\alpha_k}_\mathrm{A}\bra{{\sqrt\tau}\beta_k}. 
\eeqa

The state $\hat\rho_\mathrm{A}^\mathrm{S}$ is finally measured 
by the time-integrated homodyne detection 
to construct the Wigner function. 
As explained in Section \ref{Homodyne detection}, 
the ideal homodyne detector is sensitive not only to the state 
$\hat\rho_\mathrm{A}^\mathrm{S}$ but also 
to the vacuum states $\ket{\mathbf{0}^\mathrm{V}}$ 
outside the squeezing bandwidth. 
So the input state into the ideal homodyne detector 
must be 
\beq\label{input state into ideal homodyne detector}
\hat\rho_\mathrm{A}
=
\hat\rho_\mathrm{A}^\mathrm{S}
\otimes
\proj{\mathbf{0}^\mathrm{V}}
\eeq
The time-integrated homodyne current delivers information 
only on the LO matched mode, 
whose state is represented by the reduced density operator 
$\tilde\rho_\mathrm{A}$ 
defined by Eq. (\ref{reduced density operator}). 
Its matrix element in terms of the quadrature eigenstates is 
given by Eq. (\ref{matrix element2}). 
The Wigner function of the reduced density operator for 
the LO matched mode is then given by 
\beq\label{Wigner function:def}
W(x,p)
=
\frac{1}{2\pi}\int_{-\infty}^\infty d\xi e^{-ip\xi} 
\varbra{x+{\xi\over2}} 
  \tilde\rho_\mathrm{A} 
\varket{x-{\xi\over2}}, 
\eeq
where the matrix element is given by 
\beqa\label{matrix element for Wigner function: vector rep}
&&\varbra{x+{\xi\over2}} 
  \tilde\rho_\mathrm{A} 
\varket{x-{\xi\over2}}
\nonumber\\
&{}&
=
\int d\mathbf{x} 
\delta(x+{\xi\over2} - {}^t \bm{\epsilon} \cdot \mathbf{x}) 
\bra{\mathbf{x}}
\hat\rho_\mathrm{A} 
\ket{\mathbf{x}-\xi\bm{\epsilon}}
\nonumber\\
&{}&
=
\int d\mathbf{x}^\mathrm{S}
\int d\mathbf{x}^\mathrm{V} 
\delta( x+{\xi\over2}
       -{}^t \mathbf{w}^\mathrm{S} \cdot \mathbf{x}^\mathrm{S}
       -{}^t \mathbf{w}^\mathrm{V} \cdot \mathbf{x}^\mathrm{V})
\nonumber\\
&{}&\quad\times
\bra{\mathbf{x}^\mathrm{S}}
\hat\rho_\mathrm{A}^\mathrm{S}
\ket{\mathbf{x}^\mathrm{S}-\xi\mathbf{w}^\mathrm{S}}
\nonumber\\
&{}&\quad\times
\bra{\mathbf{x}^\mathrm{V}}
\mathbf{0}^\mathrm{V} \rangle\langle \mathbf{0}^\mathrm{V} 
\ket{\mathbf{x}^\mathrm{V}-\xi\mathbf{w}^\mathrm{V}}. 
\eeqa
In the second equality 
we have replaced the abbreviated notations 
$\{ \epsilon_k \}$ and $\{ x_k \}$ 
introduced in Eqs. (\ref{vector epsilon}) and 
(\ref{vector of eigenvalue x_phi}) 
with the explicit ones 
$\{ w_k^\mathrm{S}, w_k^\mathrm{V} \}$ 
and 
$\{ x_k^\mathrm{S}, x_k^\mathrm{V} \}$.  
Substituting the expression of Eq. (\ref{rho_A}) into 
the above equation, we have 
\beqa\label{matrix element for Wigner function: S+V space rep}
&&\varbra{x+{\xi\over2}} 
  \tilde\rho_\mathrm{A} 
\varket{x-{\xi\over2}}
\nonumber\\
&{}&
=
\frac{1}{2\pi} 
\int_{-\infty}^\infty dq e^{i(x+{\xi\over2})q}
%\Biggl[
\prod_{l=0}^\infty 
\int_{-\infty}^\infty dx_l^\mathrm{V} 
\mathrm{exp}(-iq w_l^\mathrm{V} x_l^\mathrm{V})
\nonumber\\
&{}&\quad\quad\times
\bra{x_l^\mathrm{V}}
0_l^\mathrm{V} \rangle\langle 0_l^\mathrm{V} 
\ket{x_l^\mathrm{V} - \xi w_l^\mathrm{V}}
%\Biggr]
\nonumber\\
&{}&
\times
\Biggl\{
%\Biggl[
\prod_{k=0}^\infty 
\int_{-\infty}^\infty dx_k^\mathrm{S} 
\mathrm{exp}(-iq w_k^\mathrm{S} x_k^\mathrm{S})
\nonumber\\
&{}&\quad\quad\times
\bra{x_k^\mathrm{S}}
\hat R_k(0,0)
\ket{x_k^\mathrm{S} - \xi w_k^\mathrm{S}}
%\Biggr]
\nonumber\\
&{}&\quad -
%\Biggl[
\prod_{k=0}^\infty 
\int_{-\infty}^\infty dx_k^\mathrm{S} 
\mathrm{exp}(-iq w_k^\mathrm{S} x_k^\mathrm{S})
\nonumber\\
&{}&\quad\quad\times
\bra{x_k^\mathrm{S}}
\hat R_k(\eta_k,\nu_k)
\ket{x_k^\mathrm{S} - \xi w_k^\mathrm{S}}
%\Biggr]
\Biggr\}, 
\eeqa
where we have used the Fourier expansion for the delta function 
in Eq. (\ref{matrix element for Wigner function: vector rep}). 
%
%
%
%\beqa\label{R:def}
%R(x,p;\bm{\eta},\bm{\nu}) 
%&=&
%\frac{1}{4\pi^2} 
%\int_{-\infty}^\infty d\xi e^{-ip\xi}
%\int_{-\infty}^\infty dq e^{i(x+{\xi\over2})q}
%\nonumber\\
%&\times&
%\prod_{k=0}^\infty 
%\int_{-\infty}^\infty dx_k^\mathrm{S} 
%\mathrm{exp}(-iq w_k^\mathrm{S} x_k^\mathrm{S})
%\nonumber\\
%&{}&\quad\times
%\bra{x_k^\mathrm{S}}
%\hat R_k(\eta_k,\nu_k)
%\ket{x_k^\mathrm{S} - \xi w_k^\mathrm{S}}
%\nonumber\\
%&{}&\times 
%\int_{-\infty}^\infty dx_k^\mathrm{V} 
%\mathrm{exp}(-iq w_k^\mathrm{V} x_k^\mathrm{V})
%\nonumber\\
%&{}&\quad\times
%\bra{x_k^\mathrm{V}}
%0_k^\mathrm{V} \rangle\langle 0_k^\mathrm{V} 
%\ket{x_k^\mathrm{V} - \xi w_k^\mathrm{V}}
%\eeqa
%
%
%
%\beqa\label{R:def}
%&&R(x,p;\bm{\eta},\bm{\nu}) 
%=
%\frac{1}{4\pi^2} 
%\int_{-\infty}^\infty d\xi e^{-ip\xi}
%\int_{-\infty}^\infty dq e^{i(x+{\xi\over2})q}
%\nonumber\\
%&\times&
%\prod_{k=0}^\infty 
%\int_{-\infty}^\infty dx_k^\mathrm{S} 
%e^{-iq w_k^\mathrm{S} x_k^\mathrm{S}}
%\bra{x_k^\mathrm{S}}
%\hat R_k(\eta_k,\nu_k)
%\ket{x_k^\mathrm{S} - \xi w_k^\mathrm{S}}
%\nonumber\\
%&\times& 
%\int_{-\infty}^\infty dx_k^\mathrm{V} 
%e^{-iq w_k^\mathrm{V} x_k^\mathrm{V}}
%\bra{x_k^\mathrm{V}}
%0_k^\mathrm{V} \rangle\langle 0_k^\mathrm{V} 
%\ket{x_k^\mathrm{V} - \xi w_k^\mathrm{V}}
%\eeqa
%
%
%
After straightforward calculations of Gaussian integrations, 
the Wigner function can be finally represented as 
\beq\label{Wigner function:final}
W(x,p)
=
\frac{1}{P_\mathrm{det}}
\left[ 
R(x,p;\bm{0},\bm{0})  
- 
R(x,p;\bm{\eta},\bm{\nu}) 
\right], 
\eeq
where 
\beqa\label{R:final}
R(x,p;\bm{\eta},\bm{\nu}) 
&=&
\frac{\cal{N}(\bm{\eta},\bm{\nu})}
{\pi\sqrt{\zeta_+(\bm{\eta})\zeta_-(\bm{\eta})}} 
\nonumber\\
&\times&
\exp
\Biggl[
-\frac{x^2}{\zeta_-(\bm{\eta})}
-\frac{p^2}{\zeta_+(\bm{\eta})}
\Biggr],
\eeqa
with 
\beq\label{zeta}
\zeta_\pm(\bm{\eta})
=
1\pm\tau\sum_{k=0}^\infty
\frac{2\lambda_k}{\gamma_\pm(\eta_k)}
\left( w_k^\mathrm{S} \right)^2, 
\eeq
%
%
%
%\beqa\label{zeta}
%\zeta_+(\bm{\eta})
%&=&
%1+\tau\sum_{k=0}^\infty
%\frac{2\lambda_k\epsilon_k^2}
%{\gamma_+(\eta_k)}, 
%\nonumber\\
%\zeta_-(\bm{\eta})
%&=&
%1-\tau\sum_{k=0}^\infty
%\frac{2\lambda_k\epsilon_k^2}
%{\gamma_-(\eta_k)},
%\eeqa
%
%
%
\beq\label{gamma}
\gamma_\pm(\eta_k)
=
1\mp\lambda_k\pm(1-\tau)\eta_k\lambda_k,  
\eeq
and 
\beq\label{N}
{\cal N}(\bm{\eta},\bm{\nu})
=
\prod_{k=0}^\infty
\sqrt{
\frac{1-\lambda_k^2}
{\gamma_+(\eta_k)\gamma_-(\eta_k)}
}
e^{-\nu_k},  
\eeq
and 
\beqa\label{R:0}
&&R(x,p;\bm{0},\bm{0}) 
\nonumber\\
&=&
\frac
{
\displaystyle
\frac{1}{\pi} 
\exp
\Biggl(
-\frac{x^2}{1-\tau+\sigma_-^2 \tau}
-\frac{p^2}{1-\tau+\sigma_+^2 \tau}
\Biggr)
}
{
\sqrt{(1-\tau+\sigma_-^2 \tau)(1-\tau+\sigma_+^2 \tau)}
}, 
\eeqa
%
%
%
%\beqa\label{R:0}
%R(x,p;\bm{0},\bm{0}) 
%&=&
%\frac{1}{\pi} 
%\exp
%\Biggl(
%-\frac{x^2}{1-\tau+\sigma_-^2 \tau}
%\nonumber\\
%&-&\frac{p^2}{1-\tau+\sigma_+^2 \tau}
%\Biggr)
%\nonumber\\
%&/&
%\sqrt{(1-\tau+\sigma_-^2 \tau)(1-\tau+\sigma_+^2 \tau)}, 
%\eeqa
%
%
%
with 
\beq\label{sigma_pm^2}
\sigma_\pm^2
\equiv
\sum_{k=0}^\infty 
\left[  \left( w_k^\mathrm{S} \right)^2 e^{\pm2 r_k} 
      + \left( w_k^\mathrm{V} \right)^2
\right]. 
\eeq
The probability of having the ``on" signal is given by 
\beq
P_\mathrm{det}=1-{\cal N}(\bm{\eta},\bm{\nu}). 
\eeq

Thus the relevant quantum states can be represented 
in terms of the discrete set of 
the prolate spheriodal wave function modes. 
This discretization becomes more prominent 
as the time- and band-limitation gets more stringent. 
In fact, for $BT\ll1$, only the 0th mode becomes dominant 
as shown later. 
If one could make the photon-subtracted squeezed state 
pure in this 0th single mode, 
and 
could selectively measure the 0th mode by the homodyne detector, 
then the ideal mode matching should be realized. 
But for this, it is not sufficient to simply put $BT\ll1$. 
Some additional care must be taken, 
depending on specific physical models actually used. 
So in the following two sections, 
we apply the result in this section to two kinds of models: 
CW and pulsed schemes.

%%%%%%%%%%%%%%%%%%%%%%%%%%%%%%%%%%%%%%%%%%%%%%%%%%%%%%%%%%%%%%%
\section{CW scheme}
\label{CW scheme}
%%%%%%%%%%%%%%%%%%%%%%%%%%%%%%%%%%%%%%%%%%%%%%%%%%%%%%%%%%%%%%%

%%%%%%%%%%%%%%%%%%%%%%%%%%%%%%%%%%%%%%
\begin{figure*}
\hspace{10mm}
\begin{center}
\includegraphics[width=0.7\linewidth]{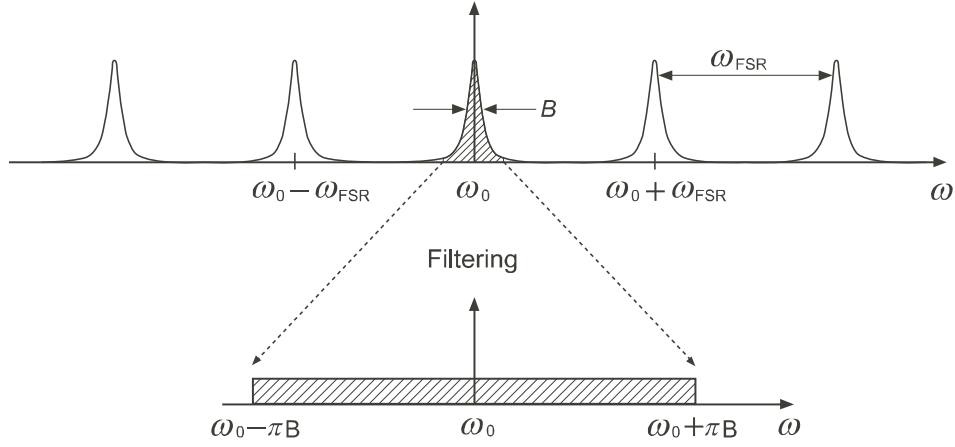}
\caption{Frequency spectrum of the squeezed state generated by 
an optical parametric oscillator cavity.}
\label{cavity_opo_spectrum}
\end{center}
\end{figure*}
%%%%%%%%%%%%%%%%%%%%%%%%%%%%%%%%%%%%%%

A typical example of CW scheme 
is an optical parametric oscillator cavity 
containing a $\chi^{(2)}$ nonlinear medium, 
continuously pumped by a single mode field 
at optical frequency $2\omega_0$. 
A pump photon is down-converted into quantum correlated twin 
photons at optical frequencies 
$\omega_0+\Omega$ and $\omega_0-\Omega$, 
resulting in an frequency entangled state of squeezed vacuum 
via cavity enhancement. 
In a commonly used temperature-controlled phase matching scheme, 
such twin photons are generated over a wide range of frequencies 
something like a few tens of GHz 
around the degenerate frequency $\omega_0$. 
A typical frequency spectrum is shown in Fig. 
\ref{cavity_opo_spectrum}. 
It consists of multi resonant peaks separated 
by the frequency $\omega_\mathrm{FSR}$ of a free spectrum range. 
For a commmon bow-tie configuration with a round-trip length of 
$\sim$500\,mm, $\omega_\mathrm{FSR}$ is about 600\,MHz, 
and a width of each resonant peak is $B\sim10$\,MHz.   
Such a wide band CW squeezed state is a source state for 
the non-Gaussian operation here. 
The events in time domain are defined 
by imposing the finite time duration $T$ on the on/off detector.

\begin{figure}
\hspace{10mm}
\begin{center}
\includegraphics[width=0.97\linewidth]{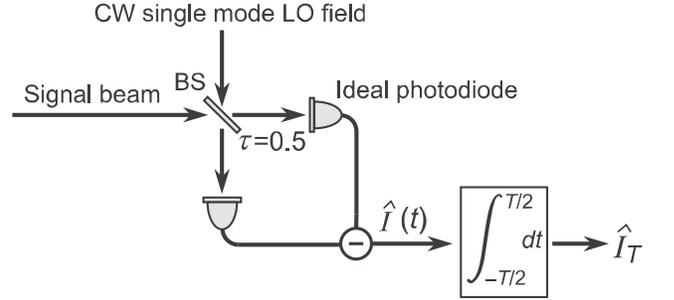}
\caption{Time-integrated homodyne detection in a CW setting. 
The detection bandwidth is assumed to be infinite.}
\label{CW_homodyne}
\end{center}
\end{figure}

Let us first consider time-integrated homodyne detection 
with the infinite detection bandwidth as shown in Fig. 
\ref{CW_homodyne}. 
The LO is a CW single mode field 
at optical frequency $\omega_0$. 
In the rotating frame, it is written as 
\beq\label{single mode LO field}
\hat A^\mathrm{L}(\Omega)
=\alpha^\mathrm{L} e^{i\phi} \delta(\Omega).   
\eeq
A CW signal beam is combined with the LO field, 
producing the CW current, 
\beq
\hat I(t)
=
\frac{1}{2\pi}\int_{-\infty}^{\infty} d\Omega 
\hat I(\Omega) e^{-i\Omega t},  
\eeq
with 
\beq
\hat I(\Omega)
=
\frac{\alpha^\mathrm{L}}{2\pi}
\left[
\hat A(\Omega)e^{-i\phi}+\hat A^\dagger(-\Omega)e^{i\phi}
\right],  
\eeq
and this current is integrated over $[-T/2, T/2]$
\beq
\hat I_{T}
=
\frac{\alpha^\mathrm{L}}{2\pi}
\int_{-\infty}^{\infty} d\Omega 
\left[
\hat A(\Omega)e^{-i\phi}
+\hat A^\dagger(\Omega)e^{i\phi}
\right]
\frac{\sin\frac{\Omega T}{2}}{\pi\Omega}.  
\eeq
As seen from this expression, 
it depends on the choice of $T$ 
what frequency modes dominate in $\hat I_{T}$. 
For $T\gg\omega_\mathrm{FSR}^{-1}$, 
contribution from the modes in the resonant peaks at 
$\omega_0\pm n\omega_\mathrm{FSR}$ $(n=1,2,...)$ 
is negligible. 
For $T>B^{-1}$, the modes inside the resonant peak 
centered at $\omega_0$ dominate. 
For an intermediate region, 
$\omega_\mathrm{FSR}^{-1}<T<B^{-1}$, 
the modes in the free spectrum range, 
which are in vacuum states, 
also contribute to $\hat I_{T}$ 
in addition to the ones in the center resonant peak.  
Thus under the assumption $T\gg\omega_\mathrm{FSR}^{-1}$, 
it is enough, 
concerning the time-integrated homodyne detection, 
to consider a squeezed state whose spectrum 
is confined to the center resonant peak shaded 
in Fig. \ref{cavity_opo_spectrum}.

%%%%%%%%%%%%%%%%%%%%%%%%%%%%%%%%%%%%%%
%%%%%%%%%%%%%%%%%%%%%%%%%%%%%%%%%%%%%%

One should, however, note that this is not true for 
the on/off detector. 
The on/off detector is sensitive to all the modes 
regardless of the measurement duration $T$. 
Therefore for a good frequency mode matching, 
the tapped-off beam  
must be optically filtered in front of the on/off detector 
such that only the center resonant peak shaded 
in Fig. \ref{cavity_opo_spectrum} 
is guided into the on/off detector. 
This can be made by employing an appropriate set 
of filtering cavities.  
This is actually the assumption 
under the model of band-limited squeezing 
introduced in Eq. (\ref{Squeezing operator}) 
or (\ref{Squeezing operator in rotating frame}). 
We further simplify the squeezing spectrum 
so as to be the flat band over $[-\pi B, \pi B]$, 
as shown in the lower part of Fig. \ref{cavity_opo_spectrum}, 
which is actually 
the model of Eq. (\ref{Flat band squeezing operator}).

%--------------------------------------------------------------
\subsection{LO matched mode measured by homodyne detector 
with the infinite bandwidth}
\label{LO matched mode measured by homodyne detector 
with the infinite bandwidth}
%--------------------------------------------------------------

Before analyzing the non-Gaussian operation, 
let us consider a simple case 
where the squeezed vacuum state 
is directly measured by the homodyne detector 
without the photon-subtraction by the on/off detector. 
The Wigner function in this case is given simply by 
$R(x,p;\bm{0},\bm{0})$ in Eq. (\ref{R:0}) and (\ref{sigma_pm^2}) 
with $\tau=1$. 
Using Eq. (\ref{definition of A_k}), 
the $k$th coefficient of the CW single mode LO field in 
Eq. (\ref{single mode LO field}) 
is given by 
\beq\label{definition of alpha_k of LO}  
\alpha_k^\mathrm{L} 
=
\frac{\alpha^\mathrm{L} }{2\pi} 
\Phi_k(c, 0)
= 
\frac{\alpha^\mathrm{L} }{2\pi} 
\sqrt{\frac{2k+1}{B}}
P_k(0), 
\eeq
where $P_k(x)$ is the $k$th Legendre polynomial and  
\beq\label{P_k(0)}
P_k(0)
=
\left\{
\begin{array}{ll}
(-1)^{\frac{k}{2}}
\displaystyle\frac{(k-1)!!}{k!!} 
& \quad (k:\mathrm{even}), 
\\
0 
& \quad (k:\mathrm{odd}). 
\end{array}\right. 
\eeq
The variance is then expressed as 
\beq\label{sigma_-^2 for single mode LO}
\sigma_-^2
=
e^{-2 \gamma} \mathrm{w}^\mathrm{S} 
+\mathrm{w}^\mathrm{V},   
\eeq
where 
\beq\label{w^S for single mode LO} 
\mathrm{w}^\mathrm{S}
\equiv
\sum_{k:\mathrm{even}} 
\left( w_k^\mathrm{S} \right)^2
=
\frac
{ \displaystyle\sum_{k:\mathrm{even}} 
  (2k+1) P_k(0)^2 \chi_k(c)^2 } 
{\displaystyle\sum_{k:\mathrm{even}} 
       (2k+1) P_k(0)^2 \chi_k(c) }, 
\eeq
and 
\beqa\label{w^V for single mode LO}
\mathrm{w}^\mathrm{V}
&\equiv&
\sum_{k:\mathrm{even}} 
\left( w_k^\mathrm{V} \right)^2
\nonumber\\
&=&
\frac
{ \displaystyle\sum_{k:\mathrm{even}} 
  (2k+1) P_k(0)^2 \chi_k(c)[1-\chi_k(c)] } 
{\displaystyle\sum_{k:\mathrm{even}} 
       (2k+1) P_k(0)^2 \chi_k(c) }. 
\eeqa

\begin{figure}
\hspace{10mm}
\begin{center}
\includegraphics[width=0.8\linewidth]{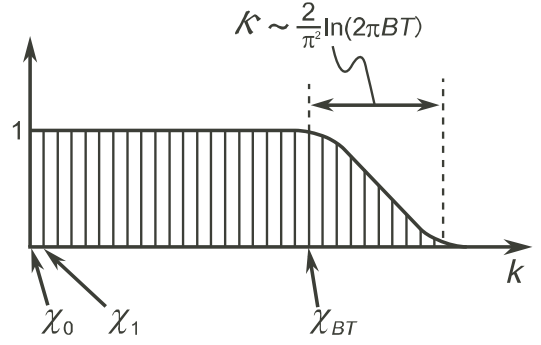}
\caption{Eigenvalu spectrum of $\chi_k(c)$ for $BT\gg1$.}
\label{eigenvalue_spectrum}
\end{center}
\end{figure}

In the case of $BT$ $(=2c/\pi)\gg1$, 
the first ${\cal{N}}\sim BT$ modes 
have eigenvalues $\chi_k(c)$ near 1. 
These modes are followed by a transition region of 
${\cal{K}}\sim 2/\pi^2\ln(2\pi BT)$ modes, 
in which the eigenvalues fall from near 1 to near 0, 
as shown in Fig. 
\ref{eigenvalue_spectrum}. 
Beyond the transition region, the remaining modes 
have eigenvalues that are very close to 0. 
This eigenvalue spectrum and the above 
Eqs. (\ref{w^S for single mode LO}) 
and (\ref{w^V for single mode LO}) 
mean that $\mathrm{w}^\mathrm{S}\sim1$ and 
$\mathrm{w}^\mathrm{V}\sim0$. 
Thus in this limit 
(a long enough measurement interval $T$) 
the original squeezing characteristics 
can be directly observed without any degradation 
due to the vacuum modes.

%%%%%%%%%%%%%%%%%%%%%%%%%%%%%%%%%%%%%%
%%%%%%%%%%%%%%%%%%%%%%%%%%%%%%%%%%%%%%

In the case of $BT\ll1$, 
on the other hand, 
the 0th eigenvalue scales as 
$\chi_0(c)\sim BT$ and $\chi_0(c)\gg \chi_k(c)$ for $k\ge1$, 
namely, only the 0th mode contributes with the weight $BT$. 
The first few eigenvalues $\chi_k(c)$ for several values of $BT$ 
are given in Table~\ref{table:eigenvalues}, 
which is borrowed from Zhu and Caves 
\cite{Zhu_Caves90}. 
This means that 
in Eq. (\ref{sigma_-^2 for single mode LO}), 
$\mathrm{w}^\mathrm{S}\sim BT\ll1$ and 
$\mathrm{w}^\mathrm{V}\sim 1-BT\approx1$. 
The second term in Eq. (\ref{sigma_-^2 for single mode LO}) 
represents the vacuum fluctuations 
due to the modes outside the squeezing bandwidth. 
Actually 
if one measures the squeezed state 
in an interval much shorter than $\sim B^{-1}$, 
one will observe wide range of frequency modes in vacuum states. 
When this term becomes dominant, 
the squeezing characteristics cannot be observed clearly, 
being covered by the vacuum fluctuation noise. 
This could be a serious restriction on optimization 
of the mode matching. 
In the next subsection we consider a scheme to remove 
this restriction by using an electrical low pass filter.

%%%%%%%%%%%%%%%%%%%%
\begin{table}
\begin{center}
\begin{tabular}{|l|@{~}l|@{~}l|@{~}l|@{~}l|}
\hline
   $k$ & $BT=0.1$ & $BT=0.5$ & $BT=1.0$ & $BT=3.0$ \\
\hline
 0 & { }0.09973 & { }0.46780 & { }0.78340 & { }0.99890 \\
 1 & { }0.00027 & { }0.03183 & { }0.20502 & { }0.96869 \\
 2 & { }0.00000 & { }0.00037 & { }0.01136 & { }0.73284 \\
 3 & { }        & { }0.00000 & { }0.00021 & { }0.26248 \\
 4 & { }        & { }        & { }0.00000 & { }0.03478 \\
 5 & { }        & { }        & { }        & { }0.00221 \\
 6 & { }        & { }        & { }        & { }0.00009 \\
\hline
\end{tabular}
\end{center}
\vspace{-1em}
\caption{\label{table:eigenvalues}
The first few eigenvalues $\chi_k(c)$ for several values of $BT$.
}
\end{table}
%%%%%%%%%%%%%%%%%%%%

%--------------------------------------------------------------
\subsection{LO matched mode measured by homodyne detector 
with a finite bandwidth}
\label{LO matched mode measured by homodyne detector 
with a finite bandwidth}
%--------------------------------------------------------------

%%%%%%%%%%%%%%%%%%%%%%%%%%%%%%%%%%%%%%
\begin{figure*}
\hspace{10mm}
\begin{center}
\includegraphics[width=0.6\linewidth]{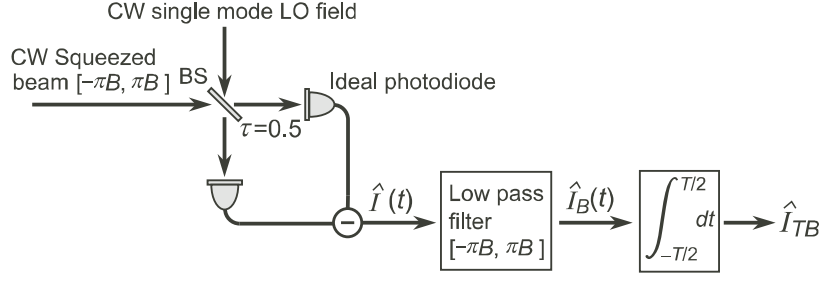}
\caption{Time-integrated homodyne detection 
with a low pass filter in a CW setting. }
\label{CW_homodyne_with_filter}
\end{center}
\end{figure*}
%%%%%%%%%%%%%%%%%%%%%%%%%%%%%%%%%%%%%%

%%%%%%%%%%%%%%%%%%%%%%%%%%%%%%%%%%%%%%
\begin{figure*}
\hspace{10mm}
\begin{center}
\includegraphics[width=0.7\linewidth]{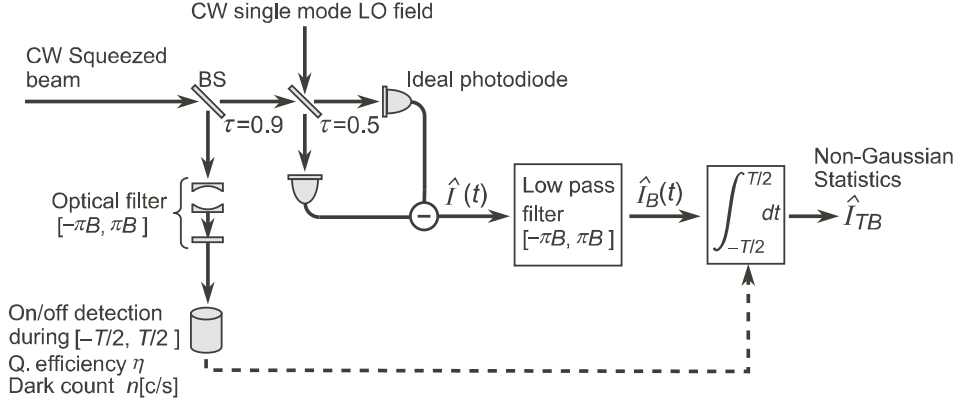}
\caption{CW scheme of the non-Gaussian operation 
by photon subtraction with the on/off detector.}
\label{CW_NonGaussian_with_filter}
\end{center}
\end{figure*}
%%%%%%%%%%%%%%%%%%%%%%%%%%%%%%%%%%%%%%

Let us consider a scheme shown in Fig. 
\ref{CW_homodyne_with_filter}, 
where an electrical low pass filter is inserted into 
the scheme of Fig. \ref{CW_homodyne}.  
The intantaneous homodyne current 
Eq. (\ref{Difference homodyne current}) 
is first filtered by the low pass filter 
with the same bandwidth $B$ matched to the squeezing spectrum, 
and is then integrated over the interval $[-T/2, T/2]$. 
The filtered current can be simply represented as 
\beq
\hat I_B(t)
=
\frac{\alpha^\mathrm{L}}{4\pi^2}\int_{-\pi B}^{\pi B} d\Omega 
\left[
\hat A(\Omega)e^{-i\phi}+\hat A^\dagger(-\Omega)e^{i\phi}
\right] 
e^{-i\Omega t}.  
\eeq
It is finally time-integrated over the interval $T$ 
\beq
\hat I_{TB}
=
\frac{\alpha^\mathrm{L}}{2\pi}
\int_{-\pi B}^{\pi B} d\Omega 
\left[
\hat A(\Omega)e^{-i\phi}
+\hat A^\dagger(\Omega)e^{i\phi}
\right]
\frac{\sin\frac{\Omega T}{2}}{\pi\Omega}.  
\eeq
In this equation the field $\hat A(\Omega)$ is band-limited 
and hence 
can be expanded in terms of $\hat A_k$ by 
Eq. (\ref{A(Omega) expansion by A_k}), 
resulting in 
\beq
\hat I_{TB}
=
\frac{\alpha^\mathrm{L}}{{\sqrt2}\pi}
\sum_{k=0}^\infty 
\hat X_k^\mathrm{S} 
\int_{-\pi B}^{\pi B} d\Omega 
\Phi_k(c,\Omega)
\frac{\sin\frac{\Omega T}{2}}{\pi\Omega},  
\eeq
where the definition 
Eq. (\ref{quadrature for squeezed mode k}) 
is used by setting $\phi=0$. 
By using the relation 
Eq. (\ref{Integral equation for Phi in text}), 
the above equation is rewritten as 
\beq\label{I_TB}
\hat I_{TB}
=
\frac{\alpha^\mathrm{L}}{{\sqrt2}\pi}
\sum_{k=0}^\infty 
\hat X_k^\mathrm{S} 
\chi_k(c)
\Phi_k(c,0).  
\eeq
So in this case the observed LO matched mode is 
specified only by the quadrature operator 
on the Hilbert space $\cal H^\mathrm{S}$ 
\beq\label{quadrature operator obtained by I_TB}
\tilde X_0
=
\sum_{k=0}^\infty 
\epsilon_k \hat X_k^\mathrm{S}, 
\eeq
where 
\beq\label{epsilon_k for homodyne detecotr with finite bandwidth}
\epsilon_k
=
\frac{\chi_k(c) \Phi_k(c,0)}
{\sqrt
{\displaystyle\sum_{k=0}^\infty \chi_k(c)^2 \Phi_k(c,0)^2}
 }, 
\eeq
which is non-zero only for even $k$, 
because of 
Eq. (\ref{definition of alpha_k of LO})
and 
Eq. (\ref{P_k(0)}). 
The observed variance Eq. (\ref{sigma_-^2 for single mode LO}) 
then reduces to 
\beq\label{sigma_-^2 for single mode LO with low pass filter}
\sigma_-^2=e^{-2 \gamma},   
\eeq
which is independent of the integration time $T$. 
Thus in this case one can observe the intrinsic characteristics 
of the squeeezing regardless of the choice of $T$. 
This is essential for achieving stringent mode matching 
in the regime of $BT \ll 1$.

%--------------------------------------------------------------
\subsection{Numerical examples and mode matching design chart}
\label{Numerical examples and mode matching design chart}
%--------------------------------------------------------------

After all, we consider the CW scheme depicted in 
Fig. \ref{CW_NonGaussian_with_filter}. 
The CW squeezed beam 
is splitted with reflectance $1-\tau=0.1$.  
The 10\% of the squeezed beam is tapped off, 
and is guided into the on/off detector, 
which is opened for the time duration $[-T/2, T/2]$ 
by electrical gating. 
This defines the discrete events in the time domain. 
The average interval between the successive trigger events 
(``on" counts) 
is assumed to be long enough compared with $T$. 
In the homodyne detector, 
the CW signal beam is combined with the CW single mode LO field, 
producing the CW current $\hat I(t)$, 
and this is filtered by the low pass filter 
matched to the squeezing bandwidth. 
Only when the ``on" signal is sent from the trigger channel, 
the filtered current $\hat I_B(t)$ is integrated over 
$[-T/2, T/2]$ synchronized with the gating signal. 
The quadrature operator of 
Eq. (\ref{quadrature operator obtained by I_TB}) 
based on the integrated current $\hat I_{TB}$ is used to 
construct the Wigner function of the conditional statistics.

As a typical model of squeezing, 
we assume $\gamma=0.35$ (3 dB squeezing) and $B=10$\,MHz. 
The net detection efficiency of the on/off detector $\eta$, 
which appears in the effective quantum efficiency for mode $k$ 
as $\eta_k=\eta\chi_k(c)$, 
is determined by the total coupling efficiency of photons 
through filters and couplers to the photondetector, 
and the intrinsic quantum efficiency of the photondetector. 
So this can be a small value something like $<$0.5. 
The mean number of dark counts $\nu_k$ for mode $k$  
can be converted into the dark count rate $n$\,[counts/s] 
by $\sum_k \nu_k=n T$. 
We vary this $n$\,[counts/s] to evaluate the dark count effect.

\begin{figure}
\hspace{10mm}
\begin{center}
\includegraphics[width=1.0\linewidth]{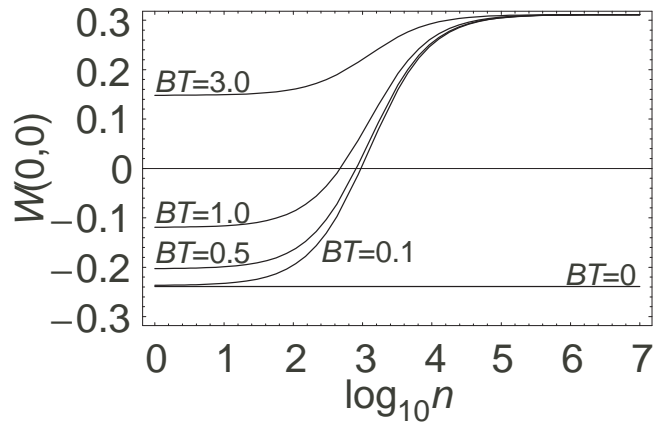}
\caption{Values of the Wigner functions at the origin of 
the phase-space, versus the dark count rate $n$\,[counts/s], 
under the condition of 3 dB squeezing, 
$\eta=0.01$, and $\tau=0.9$}
\label{W00-n_eta001}
\end{center}
\end{figure}

\begin{figure}
\hspace{10mm}
\begin{center}
\includegraphics[width=1.0\linewidth]{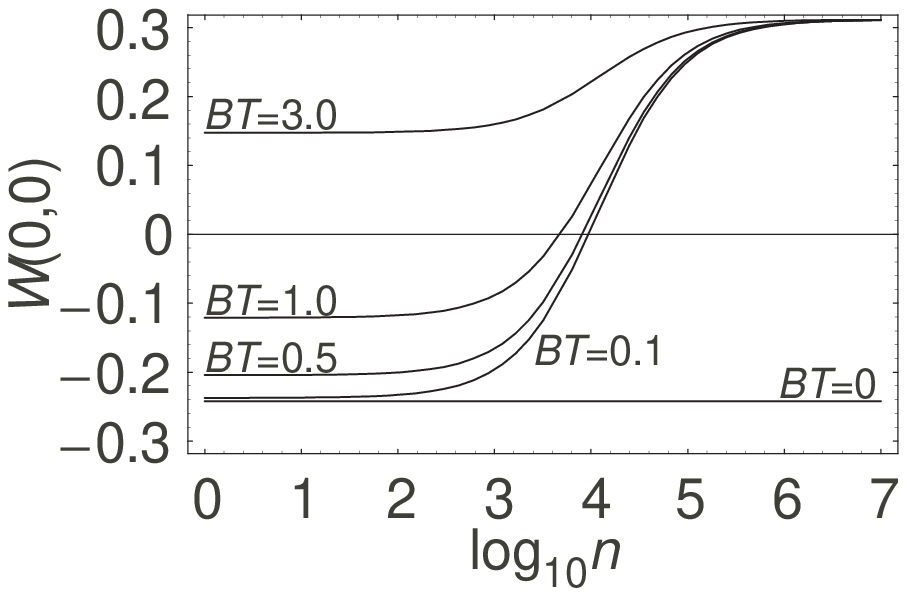}
\caption{Values of the Wigner functions at the origin of 
the phase-space, versus the dark count rate $n$\,[counts/s], 
under the condition of 3 dB squeezing, 
$\eta=0.1$, and $\tau=0.9$}
\label{W00-n_eta010}
\end{center}
\end{figure}

\begin{figure}
\hspace{10mm}
\begin{center}
\includegraphics[width=1.0\linewidth]{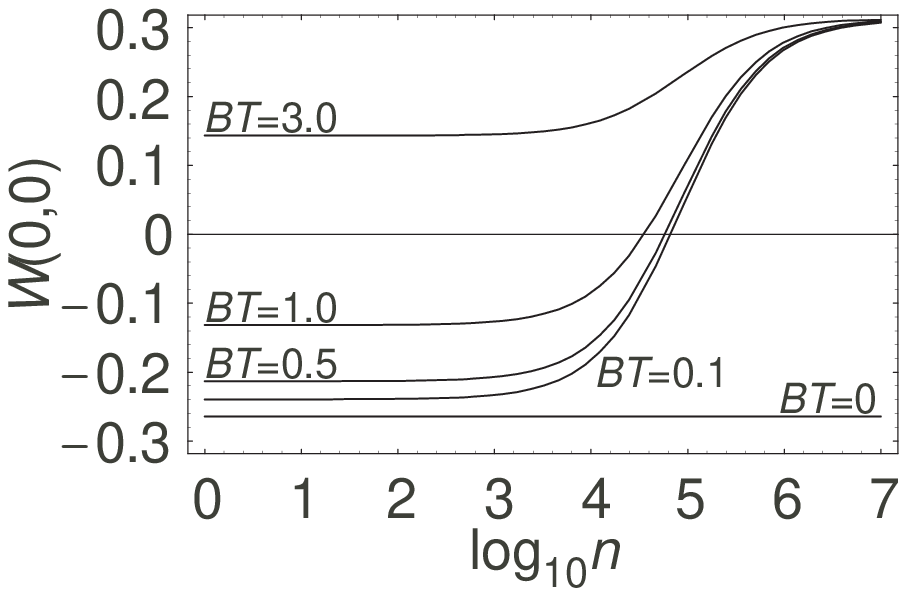}
\caption{Values of the Wigner functions at the origin of 
the phase-space, versus the dark count rate $n$\,[counts/s], 
under the condition of 3 dB squeezing, 
$\eta=0.7$, and $\tau=0.9$}
\label{W00-n_eta070}
\end{center}
\end{figure}

\begin{figure}
\hspace{10mm}
\begin{center}
\includegraphics[width=1.0\linewidth]{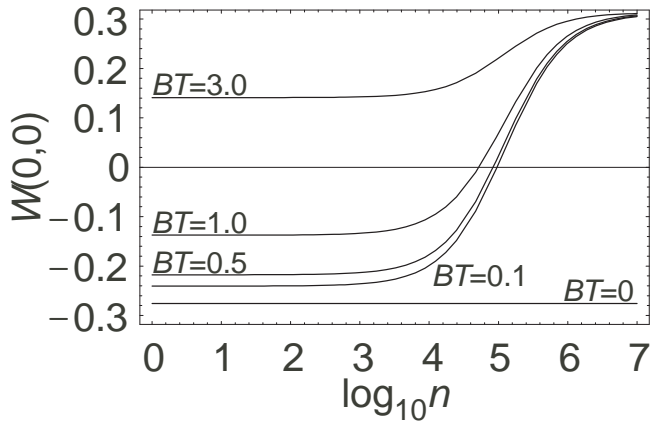}
\caption{Values of the Wigner functions at the origin of 
the phase-space, versus the dark count rate $n$\,[counts/s], 
under the condition of 3 dB squeezing, 
$\eta=1$, and $\tau=0.9$}
\label{W00-n_eta100}
\end{center}
\end{figure}

In Figs. \ref{W00-n_eta001} trough \ref{W00-n_eta100}, 
we show the values of the Wigner function at the phase space 
origin as a function of the dark count rate $n$ [counts/s] 
for several values of $BT$. 
The mode weights $\{\chi_k(c)\}$ are given in 
Table~\ref{table:eigenvalues}. 
$BT=0$ means the single mode case with perfect mode matching. 
Figs. 
\ref{W00-n_eta001}, 
\ref{W00-n_eta010}, 
\ref{W00-n_eta070}, and 
\ref{W00-n_eta100} 
corresponds to the different detection efficiencies of 
the on/off detector, $\eta=0.01$, 0.1, 0.7 and 1. 
From these figures, we can know the threshold for the on/off 
detector dark counts 
below which one can expect to observe the negative dip 
of the Wigner function, 
which is a sign of the non-classicality of the non-Gaussian 
output state. 
As the detection efficiency $\eta$ becomes smaller, 
the threshold dark count for the negative dip also gets smaller.

For larger values of $BT$, 
the mismatch between the LO matched mode and 
the photon mode observed by the on/off detector 
becomes more serious. 
For $BT=3.0$, 
one could not attain the negative dip of the Wigner function 
even with the ideal on/off detector. 
Actually, 
in the case of $BT\gg1$, 
the first ${\cal{N}}\sim BT$ modes 
have eigenvalues $\chi_k(c)$ near 1, 
and 
then the eigenvalues fall from near 1 to near 0 rapidly 
as $k$ increases, 
as shown in Fig. \ref{eigenvalue_spectrum}. 
All the first ${\cal{N}}\sim BT$ modes cause trigger 
signals at the on/off detector,  
which makes the conditional state at the signal port 
a highly mixed state, 
because the on/off detector cannot discriminate 
which mode a photon comes from. 
On the other hand, 
the homodyne detector sees only the LO matched mode, 
which is a particular combination of 
the first ${\cal{N}}\sim BT$ modes, 
and provides the homodyne statistics for 
the single mode quadrature operator 
defined by Eq. (\ref{quadrature operator obtained by I_TB}).

In the case of $BT\ll1$, on the other hand, 
$\chi_0(c)\sim BT$ and $\chi_0(c)\gg \chi_k(c)$ for $k\ge1$. 
So only the 0th mode is dominant 
both in the trigger channel and 
the homodyne detector: 
more precisely, 
the POVM element for the ``on" signal  
$\hat{\Pi}_\mathrm{on}^\mathrm{B}(\bm{\eta},\bm{\nu})$ 
in Eq. (\ref{on/off POVM}) 
and 
the quadrature eigenstate $\varket{x}$ 
on the Hibert space ${\cal H}^\mathrm{L}$ 
describing the homodyne detector 
(see the text from Eq. (\ref{tilde eigenstate of mode 0}) 
to Eq. (\ref{matrix element2})).   
If one further takes a small reflectance for 
the tapping beam splitter, 
making the probability of detecting more than two photons 
at the on/off detector very small, 
then the trigger photons are projected onto 
an almost pure single photon state. 
The homodyne detector measures 
the same pure state with almost perfect efficiency, 
attaining the best mode matching. 
The small tapping fraction, 
combined with a small value of $\chi_0(c)\sim BT$, 
means a small efficiency for the on/off detector. 
This is usually unwanted, 
but in the present context this simply results in 
the reduction of the number of selected events. 
If the number of true trigger events can be kept larger than 
the dark counts, 
this reduction would be an acceptable sacriface 
to attain better mode matching.

One should note that it is essential to filter 
the homodyne current by the low pass filter 
matched to the squeezing bandwidth $B$ 
before integrating over the interval $[-T/2, T/2]$. 
If the ideal homodyne detector with wider enough bandwidth 
than $B$ would be used, 
the vacuum fluctuation would also become dominant 
in the homodyne statistics especially for shorter $T$. 
This component does not have any mode overlap with 
the trigger photons at the on/off detector, and 
would degrade the mode matching quality.

\begin{figure}
\hspace{10mm}
\begin{center}
\includegraphics[width=1.0\linewidth]{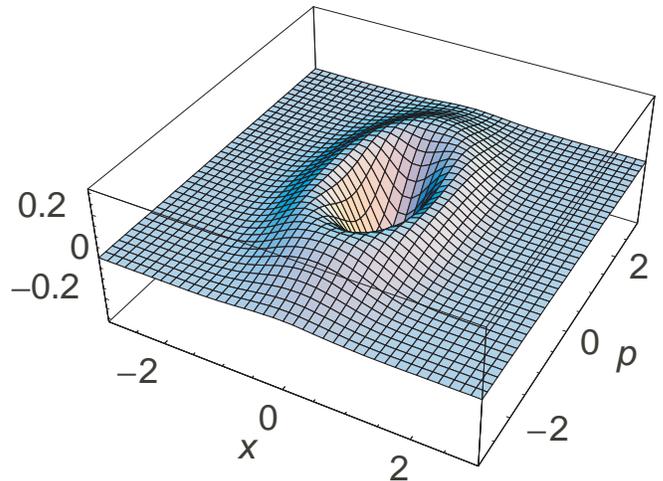}
\caption{Wigner function distribution for $BT=0$, i.e. 
the ideal single mode case, at the point of dark count 
$n=500$\,[counts/s] 
in Fig. \ref{W00-n_eta010} 
(3 dB squeezing, $\tau=0.9$, $\eta=0.1$)}
\label{WF_eta010_500cps_BT00}
\end{center}
\end{figure}

\begin{figure}
\hspace{10mm}
\begin{center}
\includegraphics[width=1.0\linewidth]{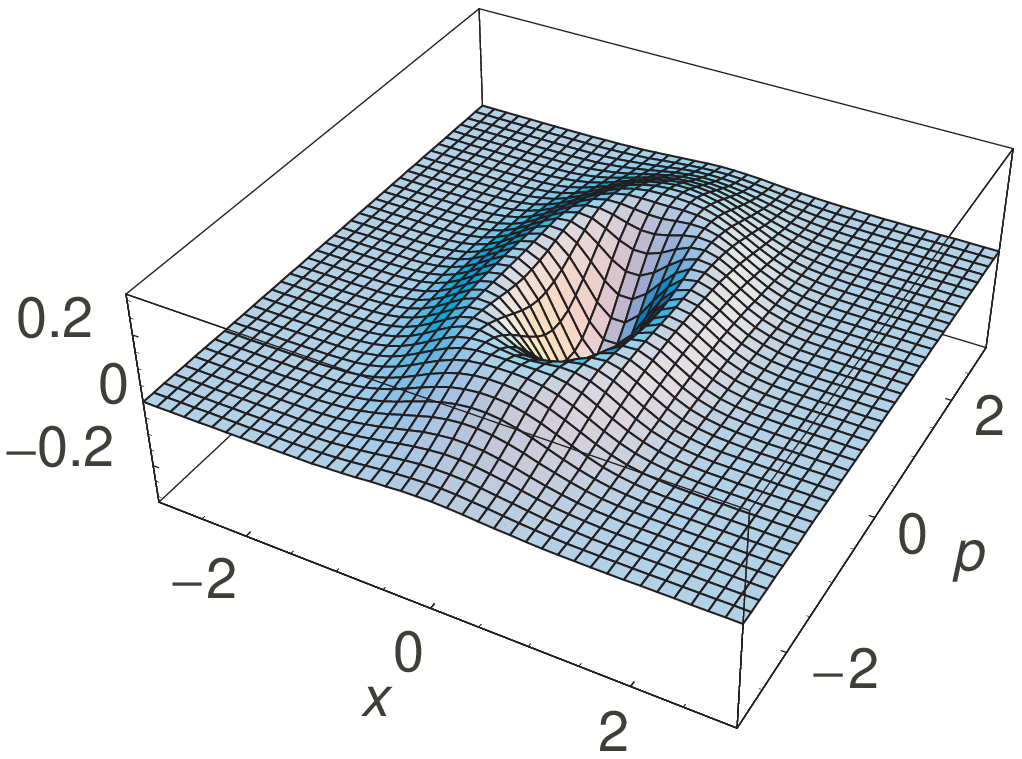}
\caption{Wigner function distribution for $BT=0.5$ 
at the point of dark count $n=500$\,[counts/s] 
in Fig. \ref{W00-n_eta010} 
(3 dB squeezing, $\tau=0.9$, $\eta=0.1$)}
\label{WF_eta010_500cps_BT05}
\end{center}
\end{figure}

\begin{figure}
\hspace{10mm}
\begin{center}
\includegraphics[width=1.0\linewidth]{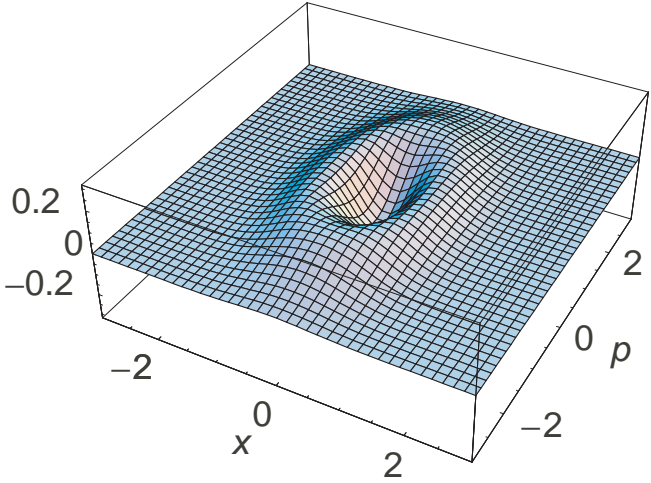}
\caption{Wigner function distribution for $BT=1.0$ 
at the point of dark count $n=500$\,[counts/s] 
in Fig. \ref{W00-n_eta010} 
(3 dB squeezing, $\tau=0.9$, $\eta=0.1$)}
\label{WF_eta010_500cps_BT10}
\end{center}
\end{figure}

Figs. \ref{WF_eta010_500cps_BT00}, 
\ref{WF_eta010_500cps_BT05}, and 
\ref{WF_eta010_500cps_BT10} 
represent the Wigner function distributions 
at the point of dark count $n=500$\,[counts/s] 
in Fig. \ref{W00-n_eta010} ($\eta=0.1$) 
for the three kinds of $BT=$ 0.0, 0.5, and 1.0, 
where one can still expect the negative dip.

%%
%%

%%%%%%%%%%%%%%%%%%%%%%%%%%%%%%%%%%%%%%%%%%%%%%%%%%%%%%%%%%%%%%%
\section{Pulsed scheme}
\label{Pulsed scheme}
%%%%%%%%%%%%%%%%%%%%%%%%%%%%%%%%%%%%%%%%%%%%%%%%%%%%%%%%%%%%%%%

In this section 
we consider a scheme 
using Fourier-transform limited short pulses. 
A pulse width of the fundamental field (LO field) 
is typically 1\,ps or shorter, 
so that the spectral width is about 1\, THz or wider. 
The pump field is the frequency-doubled LO 
with almost the same spectral width. 
A nonlinear crystal is pumped in a traveling-wave configuration 
in a single path. 
Then unlike the CW configuration, 
the down-converted (squeezed) field has generally 
different mode properties 
from the ones of the pump and the LO fields. 
The finiteness of the nonlinear crystal broadens 
the phase-matching condition, 
producing a wider bandwidth of squeezing than that of the LO. 
The group velocity mismatch increases the pulse duration, 
and makes the field apart from the Fourier-transform limit. 
The group velocity dispersion induces a frequency charp. 
These discrepancies 
between the signal and the LO field properties 
degrade the degree of mode matching. 
In order to overcome this problem, 
the usage of optical filters with narrow bandwidth 
in the trigger channel was enphasized in 
\cite{Ou97QSemiclOpt,Grosshans01,Aichele Lvovsky Schiller 02EurPhysJD,Viciani Zavatta Bellini 04PRA}. 
In these works, 
a biphoton source state  
\beqa
\ket{\rho}_\mathrm{AB}
=
\ket{0}_\mathrm{A} \ket{0}_\mathrm{B}
-i \int d^3 k d\omega d^3 k' d\omega' 
\nonumber\\
\times
\Psi(\mathbf{k},\omega,\mathbf{k}',\omega')
\ket{\mathbf{k},\omega}_\mathrm{A} 
\ket{\mathbf{k}',\omega'}_\mathrm{B}
\eeqa
is assumed. 
Here $\ket{\mathbf{k},\omega}_\mathrm{j}$ means 
the single photon state with momentum $\mathbf{k}$ 
and frequency $\omega$ in path j (=A, B), 
and the function 
$\Psi(\mathbf{k},\omega,\mathbf{k}',\omega')$ 
characterizes the correlation properties of the biphoton state. 
Since we are concerned here with frequency mode matching, 
we shall suppress the momentum representation. 
The trigger photon at path B is selected by frequency filters, 
and is then detected by an ideal photon counter. 
When the photon counter clicks, 
the trigger state is projected onto a POVM element 
\beq\label{POVM element for on signal:Aichele et al}
\hat\Pi_\mathrm{on} 
=
\int d\omega F(\omega) 
\ket{\omega}_\mathrm{B} \bra{\omega}, 
\eeq
where $F(\omega)$ includes filter characteristics and 
the effective quantum efficiency of the photon counter 
for each frequency mode. 
The conditional signal state at path A is then given by 
\beq
\hat\rho_\mathrm{A}
=
\frac
{\mathrm{Tr}_\mathrm{B}
 \left(
    \hat\Pi_\mathrm{on} 
    \ket{\rho}_\mathrm{(AB)} \bra{\rho}
 \right)}
{\mathrm{Tr}_\mathrm{AB}
 \left(
    \hat\Pi_\mathrm{on} 
    \ket{\rho}_\mathrm{(AB)} \bra{\rho} 
 \right)}.
\eeq
In \cite{Aichele Lvovsky Schiller 02EurPhysJD}, 
the correlation function 
\beq
\Gamma(\omega, \omega')
=
\mathrm{Tr}
\left[
  \hat\rho_\mathrm{A} 
  \hat a^\dagger(\omega) \hat a(\omega)
\right], 
\eeq
and the modal purity 
\beq
P
=
\mathrm{Tr}
\left(
  \hat\rho_\mathrm{A}^2
\right), 
\eeq
were introduced, 
and a measure of mode matching $M$ is defined 
using the correlation function. 
The quality of mode matching was analyzed 
in terms of these $M$ and $P$. 
However, 
it has been unclear how to obtain explicit expressions 
of the quantum states when these quantities are given.

Our interest here is a more direct quantity than $M$ and $P$, 
that is, 
the Wigner function of $\hat\rho_\mathrm{A}$ 
finally obtained by the homodyne detection. 
In addition, 
the source state is not the photon pair but 
the wideband squeezed state, 
which is beyond the first order perturbation theory. 
The source state is a flat band squeezed state 
$\ket{\mathbf{r}_0}=\hat S_0\ket{\mathbf0}$ 
\beqa\label{Squeezing operator in rotating frame:Pulsed scheme}  
\hat S_0
=\mathrm{exp}
\Biggl(
\frac{\gamma}{4\pi}
\int_{-\pi B_\mathrm{S}}^{\pi B_\mathrm{S}} d\Omega
\biggl[
\hat A(\Omega)\hat A(-\Omega)
\nonumber\\
-
\hat A^\dagger(\Omega)\hat A^\dagger(-\Omega)
\biggr]
\Biggr), 
\eeqa
where the squeezing bandwidth $B_\mathrm{S}$ 
is usually much wider than that of the LO $B$, 
i.e. $B_\mathrm{S}\gg B$. 
Here we have omitted the group velocity dispersion and mismatch, 
because this simply causes slight quantitative modification 
for the filtered state, 
and is not essential in considering 
the frequency mode matching here.

We should first derive the quadrature eigenstate 
$\varket{x}$ to describe the POVM of the homodyne detector. 
For simplicity, 
we assume the LO field with a flat band spectrum 
over the bandwidth $B$ 
\beq\label{DFT functions}
\hat A^\mathrm{L}(\Omega)
=
\left\{
\begin{array}{ll}
\alpha^\mathrm{L}
e^{i\phi} 
& \quad 
 (\vert\Omega\vert \le \pi B), 
\\
0 
& \quad 
\mathrm{otherwise}. 
\end{array}\right. 
\eeq
The instantaneous homodyne current is then given by 
\beq
\hat I(t)
=
\frac{1}{2\pi}\int_{-\infty}^{\infty} d\Omega 
\hat I(\Omega) e^{-i\Omega t}
\eeq
with 
\beqa
\hat I(\Omega)
&=&
\frac{1}{2\pi}
\int_{-\infty}^{\infty} d\Omega'
\biggl[
  \hat A^\mathrm{L\dagger}(\Omega') \hat A(\Omega+\Omega')
\nonumber\\
&{}&
+\hat A^\dagger(-\Omega+\Omega')\hat A^\mathrm{L}(\Omega') 
\biggr]
\nonumber\\
&=&
\frac{\alpha^\mathrm{L}}{2\pi}
\int_{-\pi B}^{\pi B} d\Omega'
\biggl[
  \hat A(\Omega+\Omega')e^{-i\phi}
\nonumber\\
&{}&
 +\hat A^\dagger(-\Omega+\Omega')e^{i\phi}
\biggr]. 
\eeqa
Note that the field operator 
at the signal port $\hat A(\Omega)$ itself 
is not band-limited. 
If the homodyne detector would have the infinite bandwidth, 
then the instantaneous current $\hat I(t)$ 
provides the final observable. 
In practice, however, this is not the case. 
Rather, the effective bandwidth of the homodyne detector 
including the photodiodes and the electronics, $B_\mathrm{H}$ 
is much narrower than that of the LO. 
Typically, $B_\mathrm{H}$ is a few hundred MHz at most 
while $B\ge$1\,THz. 
So one actually measures the low frequency components 
within the homodyne detector bandwidth. 
We shall approximately model it as 
\beqa
\hat I_{B_\mathrm{H}}(t)
&=&
\frac{1}{2\pi}\int_{-\pi B_\mathrm{H}}^{\pi B_\mathrm{H}} d\Omega 
\hat I(\Omega) e^{-i\Omega t}
\nonumber\\
&=&
\frac{\alpha^\mathrm{L}}{4\pi^2}
\int_{-\pi B_\mathrm{H}}^{\pi B_\mathrm{H}} d\Omega
e^{-i\Omega t}
\int_{-\pi B}^{\pi B} d\Omega'
\nonumber\\
&{}&\times
\biggl[
  \hat A(\Omega+\Omega')e^{-i\phi}
 +\hat A^\dagger(-\Omega+\Omega')e^{i\phi}
\biggr]
\nonumber\\
&\approx&
\frac{\alpha^\mathrm{L}}{2\pi}
\cdot
\frac{\sin\pi B_\mathrm{H}t}{\pi t}
\nonumber\\
&{}&\times
\int_{-\pi B}^{\pi B} d\Omega
\biggl[
  \hat A(\Omega)e^{-i\phi}
 +\hat A^\dagger(\Omega)e^{i\phi}
\biggr]. 
\eeqa
The field operator $\hat A(\Omega)$ is now band-limited to $B$, 
and can be expanded by the discrete set of 
the prolate spheroidal wave function basis.  
Using the formula in Section \ref{Time and band-limited signals}, 
we have 
\beqa
\hat I_{B_\mathrm{H}}(t)
\approx
{\sqrt2}\alpha^\mathrm{L}
\frac{\sin\pi B_\mathrm{H}t}{\pi t}
\sum_{k=0}^\infty \hat X_k^\mathrm{S}(\phi) \Psi_k(c, 0), 
\eeqa
where the quadrature operator $\hat X_k^\mathrm{S}(\phi)$ 
is defined by Eq. (\ref{quadrature for squeezed mode k}). 
The time dependent factor means that 
a short optical pulse with a duration $T\sim B^{-1}$, 
say $\sim1$\, ps, 
is converted into an electrical pulse 
with a duration $T_\mathrm{H}\sim B_\mathrm{H}^{-1}\sim0.1$\, ns. 
As seen from this equation 
it is sufficient to sample the peak value of the electrical pulse 
\beqa
\hat I_{B_\mathrm{H}}(0)
\approx
{\sqrt2}\alpha^\mathrm{L}B_\mathrm{H}
\sum_{k=0}^\infty \hat X_k^\mathrm{S}(\phi) \Psi_k(c, 0), 
\eeqa
for the quadrature values. 
The final observable is then given by 
\beq\label{quadrature operator in pulsed scheme}
\tilde X_0
=
\sum_{k=0}^\infty 
\epsilon_k \hat X_k^\mathrm{S}, 
\eeq
where 
\beqa\label{epsilon_k for pulsed homodyne detecotr}
\epsilon_k
&=&
\frac{\Psi_k(c,0)}
{\sqrt
{\displaystyle\sum_{k=0}^\infty \vert\Psi_k(c,0)\vert^2}
 }, 
\nonumber\\
&=&
\frac{(-1)^k\sqrt{(2k+1)\chi_k(c)}P_k(c,0)}
{\sqrt
{\displaystyle\sum_{k=0}^\infty (2k+1)\chi_k(c)P_k(c,0)^2}
 }, 
\eeqa
which is non-zero only for even $k$. 
The POVM of the pulsed homodyne detector is represented 
by the projection $\varket{x}\varbra{x}$ 
onto the eigenstates of $\tilde X_0$, 
tracing out all the other modes.

If the trigger beam is projected onto the single photon state 
$\varket1$ on the subspace $\cal H^\mathrm{L}$ 
spanned by $\varket{x}$, 
the ideal photon-subtracted non-Gaussian state would be obtained.  
Toward this limit, 
the trigger beam must be band-limited to $B$, 
since the homodyne observable consists of the frequency modes 
within the LO bandwidth $\vert\Omega\vert\le\pi B$. 
It can be made by interference band pass filters 
typically with 0.1-1\,nm spectral width. 
The band-limited trigger beam is finally detected 
by the on/off detector. 
The on/off detector responds much more slowly, 
say in a scale of $T_\mathrm{D}\agt300$\,ps, 
than the pulse width $T\alt1$\,ps. 
But if we assume that 
the average interval between ``on" signals 
is still much longer than this $T_\mathrm{D}$, 
and also 
that the photoelectric conversion does not significatly destroy 
the Fourier-transform limited pulses. 
Then the on/off detector can be represented by the state basis 
of the prolate spheroidal wave functions for the $BT$, 
as described in Section \ref{On/off detector}.

The photon subtracted non-Gaussian state at path A is given by 
\beq\label{photon subtracted state:Pulsed scheme}
\hat\rho_\mathrm{A}
=
\frac{
\tr_\mathrm{B}
\left[ 
\ket{\rho}_\mathrm{(AB)}\bra{\rho}
\hat I^\mathrm{A}\otimes\hat{\Pi}_\mathrm{on}^\mathrm{B}
(\bm{\eta},\bm{\nu})
\right]
}
{
\tr_\mathrm{AB}
\left[ 
\ket{\rho}_\mathrm{(AB)}\bra{\rho}
\hat I^\mathrm{A}\otimes\hat{\Pi}_\mathrm{on}^\mathrm{B}
(\bm{\eta},\bm{\nu})
\right]
}.  
\eeq
Since $\hat{\Pi}_\mathrm{on}^\mathrm{B}$ 
detects only the modes in $\vert\Omega\vert\le\pi B$, 
and these modes are not quantum correlated 
with the ones outside of it, 
the state $\ket{\rho}_\mathrm{AB}$ can be regarded as 
the beam splitted state from the squeezed state 
$\ket{\mathbf{r}}=\hat S\ket{\mathbf0}$ where 
\beqa\label{Filtered squeezing operator:Pulsed scheme}  
\hat S
&=&
\mathrm{exp}
\Biggl(
\frac{\gamma}{4\pi}
\int_{-\pi B}^{\pi B} d\Omega
\biggl[
\hat A(\Omega)\hat A(-\Omega)
\nonumber\\
&{}&-
\hat A^\dagger(\Omega)\hat A^\dagger(-\Omega)
\biggr]
\Biggr)
\nonumber\\
&=&
\bigotimes_{k=0}^{\infty}
\mathrm{exp}
\left[
  \frac{r_k}{2}
  \left( \hat A_k^2 - \hat A_k^{\dagger2} \right)
\right], 
\eeqa
instead of the squeezed state 
$\ket{\mathbf{r}_0}=\hat S_0\ket{\mathbf0}$ 
of 
Eq. (\ref{Squeezing operator in rotating frame:Pulsed scheme}). 
Thus the photon subtracted state $\hat\rho_\mathrm{A}$ 
is represented in terms of 
$\{\hat A_k\}$ or $\{\hat X_k^\mathrm{S}\}$ 
and 
its eigenstates $\{\ket{x_k^\mathrm{S}}\}$. 
Its Wigner function can then be calculated 
by the formulus (\ref{Wigner function:final}).
The mode matching consideration proceeds just as 
the CW case in 
Section \ref{Numerical examples and mode matching design chart}. 
Good mode matching is achieved 
when the weight coefficient $\epsilon_0$ dominates the other 
ones $\epsilon_1, \epsilon_2,...$. 
This can be made by setting the valuse $BT$ as small as possible. 
In the pulsed scheme, however, 
the $T$ is automatically set by the laser pulse width. 
It is actually impossible at present 
to realize a shorter measurement interval 
because electrical gating cannot reach such a time scale. 
Therefore the value of $BT$ is lower bounded 
by the Fourier-transform limit, 
which is about $BT\approx1$ or slightly less. 
Then a few lower modes can contribute, 
as shown in Table~\ref{table:eigenvalues}. 
This is already enough to observe the negative dip 
in the Wigner function, 
provided that mode matching with respect to 
the other degrees of freedom can be made perfect.

In experimental practice of the pulsed scheme, however, 
it is generally not easy to realize 
a high quality spatiotemporal mode matching 
unlike the CW scheme based on cavity OPO systems. 
For meaningful numerical evaluations, 
consideration on imperfections due to the spatial mode mismatch 
should be involved. 
But this is beyond the scope of this paper.

%%%%%%%%%%%%%%%%%%%%%%%%%%%%%%%%%%%%%%%%%%%%%%%%%%%%%%%%%%%%%%%
\section{Conclusion}
\label{Conclusion}
%%%%%%%%%%%%%%%%%%%%%%%%%%%%%%%%%%%%%%%%%%%%%%%%%%%%%%%%%%%%%%%

We have developed a theory to design the frequency mode matching 
in the photon-subtracting operation 
by the on/off-type photon detector 
on the wideband squeezed state. 
It is essential to represent the POVMs 
of the on/off photon detection and the homodyne detection 
in terms of an appropriate basis 
under time- and band-limitation on the fields. 
Such a set is based on 
the prolate spheroidal wave functions 
which are complete and orthonormal 
under the time- and band-limitation. 
The POVMs thus represented define the measured modes, 
and hence 
the quantum states of the trigger photons and 
the conditionally selected homodyne events. 
The mode matching is pursued for those quantum states.

Our theory has been applied to the CW and pulsed schemes. 
In both schemes, 
the desired condition for good frequency mode mathching 
is $BT\alt1$. 
In the previous works 
\cite{Ou97QSemiclOpt,Grosshans01,Aichele Lvovsky Schiller 02EurPhysJD,Viciani Zavatta Bellini 04PRA}, 
only narrowband spectral filtering is emphasized. 
But the quantity to be set smaller is the product $BT$. 
In addition, 
the modes of interest are not the ones specified 
by the plane wave basis
~\cite{Ou97QSemiclOpt,Grosshans01,Aichele Lvovsky Schiller 02EurPhysJD,Viciani Zavatta Bellini 04PRA} 
but rather the ones specified 
by the prolate spheroidal wave functions 
under the time- and band-limitation.  
In the regime $BT\alt1$, 
the discreteness of modes becomes prominant, 
and only a few lower modes 
of the prolate spheroidal wave functions are excited.  
For $BT\alt0.5$, only the 0th mode becomes dominant 
with the mode weight $\chi_0(c)\sim BT$ 
as seen in Table~\ref{table:eigenvalues}. 
Then the trigger channel selects a pure single photon state 
in mode 0 
if one takes a small reflectance $1-\tau$ of 
the tapping beam splitter, and could suppress dark counts. 
The homodyne detector measures 
the same pure state with almost perfect efficiency, 
attaining the ideal mode matching.

The increase of $BT$ and the dark counts 
makes the conditional state more mixed over many modes of $k$, 
while the homodyne detector still projects the conditional state 
onto a pure eigenstate of the LO matched quadrature operator, 
which is a particular combination of 
the first ${\cal{N}}\sim BT$ modes. 
Then the non-classical effect in homodyne statistics 
will be smeared out.

In the CW scheme 
where all the beams are CW, 
such as based on cavity OPO systems, 
the effective bandwidth can be set 
in a range of $B\approx10$\, MHz 
by using an appropriate set of filtering cavities 
placed in fromt of the on/off detector, 
and an electrical low pass filter in the homodyne detector 
spectrally matched to the squeezing. 
Then the time duration $T$ for the desired condition $BT\alt1$ 
is of order of sub $\mu$s, 
which can readily be achieved by current electrical gating. 
A smaller $T$ allows one more stringent frequency mode matching. 
The lower bound for $T$ may be set by the temporal resolution 
of the homodyne detector, 
which is currently of order of a few ns. 
So the frequency mode matching in a regime $BT\alt0.1$ 
can be attained in principle. 
The spatial mode matching can also be fulfilled 
by carefully locking the cavities. 
The temporal mode matching in the sub $\mu$s is not a problem. 
Then the numerical results presented in Section
~\ref{Numerical examples and mode matching design chart} 
can be practical design charts. 
One may expect to realize the negative dip of the Wigner 
function at the phase space origin for practical 
experiental parameters. 
All in all, 
the CW scheme will provide a good test-bed 
for the non-Gaussian operations 
based on photon counting and homodyning.

In the pulsed scheme, 
the time duration $T$ is automatically set 
by the laser pulse width, typically ps order. 
The band limitation comes from the LO bandwidth, 
$\approx1$\,THz, 
corresponding to the Fourier-transform limit 
$0.5\alt BT\alt1$. 
The time response of the homodyne detector is much slower 
than the LO pulse width. 
Then by sampling the peak value of the homodyne current pulse, 
one finally observes the quadrature 
composed linearly of a few lower modes of $k$. 
This makes the situation very similar to the CW case, 
and the numerical examples in Section 
\ref{Numerical examples and mode matching design chart} 
can apply. 
So the frequency mode matching needed 
for observing the negative dip of the Wigner function 
will be possible. 
In practical experiments, such as ref. 
\cite{Wenger04}, 
however, 
the observed dip of the Wigner function is still positive. 
This may be partly attributed to the imperfect spatial mode 
mathing between the trigger photons and the LO field. 
It is generally more difficult, 
compared with the cavity CW scheme, 
to control the wave front of the pulsed field, 
and to spatially match the relevant modes.

Now we would like to mention the relation between 
our result and the one obtained by Grosshans and Grangier 
\cite{Grosshans01}. 
They predicted the lower degree of mode matching 
for the CW scheme than for the pulsed one. 
This seems to be contradict with our result. 
In \cite{Grosshans01}, however, some points are missing. 
Firstly, 
the POVM element in the trigger channel was modeled by 
a projection onto a pure state 
\beq\label{POVM element for on signal:Grosshans et al}
\hat\Pi_\mathrm{on} 
=
\int_{-\pi B}^{\pi B} d\Omega  
\ket{\Omega} 
\int_{-\pi B}^{\pi B} d\Omega' 
\bra{\Omega'}, 
\eeq
instead of 
Eq. (\ref{POVM element for on signal:Aichele et al}). 
But photon counters cannnot be sensitive to the relative phases 
of different frequency components. 
So the modeling by the above equation is unlikely, 
as also pointed out by Aichele  et al. 
\cite{Aichele Lvovsky Schiller 02EurPhysJD}. 
Secondly, 
the parameter $X=\delta\omega_i\delta T/2\pi$ 
in \cite{Grosshans01}, 
which corresponds to $BT$ in this paper, 
could not be arbitrary small 
for the Fourier-transform limited pulses. 
According to these points, 
the degree of the mode overlap $\eta_\mathrm{eff}(X)$ 
should be reconsidered in the pulsed regime. 
Finally, 
the analysis on the ``chopped" CW scheme 
did not take the time- and band-limitation into acount. 
In \cite{Grosshans01}, 
the single parameter 
\beq
\eta_\mathrm{eff}(BT)
=
\frac{1}{BT}
\left\vert
\int_{-\pi B}^{\pi B} d\Omega 
\frac{\sin\frac{\Omega T}{2}}{\pi\Omega}
\right\vert^2,  
\eeq
was used for quantifying the degree of mode matching. 
But this cannot be sufficient 
for treating the time- and band-limited signals. 
More precisely the mode matching must be characterized 
by the set of mode weights 
\beq
\int_{-\pi B}^{\pi B} d\Omega 
\Phi_k(c,\Omega)
\frac{\sin\frac{\Omega T}{2}}{\pi\Omega}
=\chi_k(c) \Phi_k(c,0).  
\eeq
Therefore the upper bound $\eta_\mathrm{eff}(X)\le0.825$ 
for the CW scheme derived in \cite{Grosshans01} 
will not be the true limit. 
Rather the more stringent matching will be possible 
as shown in this paper.

Concerning to the CW scheme, 
we should also mention the effect of chopping the beams 
for the duration $T$. 
The scheme in this paper is to use the CW single-mode LO 
and then to integrate the CW homodyne current over $T$. 
This is equivalent to 
using the chopped LO with the duration $T$, 
if the detector bandwidth is infinite. 
Remember that the homodyne detector with the infinite bandwidth 
suffers from the vacuum fluctuations 
outside the squeezing bandwidth for shorter $T$. 
So we have considered installing the matched low pass filter. 
In such a band-limited case, 
using the chopped LO will not generally be equivalent to 
integrating the CW current over $T$ afterward. 
In fact, 
chopping the CW single-mode LO modifies its spectrum from 
$\delta(\Omega)$ to $\sin(\Omega T/2)/\pi\Omega$. 
Then the homodyne current is also modulated accordingly, 
and is then filtered. 
According to the present analysis, 
the use of the chopped LO does not seem 
to bring any particular merit. 
Therefore it will be enough to simply use the CW single-mode LO, 
and 
to electrically gate the on/of detetor to define the events in  
the time domain.

\begin{acknowledgments}
The authors would like to thank K.~Wakui, Y.~Takahashi, 
M.~Takeoka, K.~Tsujino, P.~Kumar, A.~I.~Lvovsky, 
and A.~Furusawa 
for valuable discussions. 
\end{acknowledgments}

%%%%%%%%%%%%%%%%%%%%%%%%%%%%%%%%%%%%%%%%%%%%%%%%%%%%%%%%%%%%%%%

\appendix*

%%%%%%%%%%%%%%%%%%%%%%%%%%%%%%%%%%%%%%%%%%%%%%%%%%%%%%%%%%%%%%%
\section{Prolate Spheroidal Wave Functions}
\label{Appendix:Prolate Spheroidal Wave Functions}
%%%%%%%%%%%%%%%%%%%%%%%%%%%%%%%%%%%%%%%%%%%%%%%%%%%%%%%%%%%%%%%

In this appendix basic properties of the prolate spheroidal 
wave functions are sumarized for reader's convenience. 
The differential equation 
\beq\label{differential equation}
(1-x^2)\frac{d^2u}{dx^2}
-2x\frac{du}{dx}
+(\mu-c^2 x^2)u=0, 
\eeq
has continuous solutions in the closed set $x$ interval 
$[-1,1]$ only for certain discrete real positive values, 
\beq
0<\mu_0(c)<\mu_1(c)<\mu_2(c)<\cdots. 
\eeq
Corresponding to each eigenvalue $\mu_k(c)$, there is a unique 
solution $S_{0k}(c,x)$ 
such that it reduces to the $k$th Legendre polynomial $P_k(x)$ 
uniformly in $[-1,1]$ as $c\rightarrow0$. 
The functions $S_{0k}(c,x)$ are called 
the angular prolate spheroidal wave functions. 
They are real for real $x$, 
continuous functions of $c$ for $c\ge0$, and 
orthogonal in $(-1,1)$.  
$S_{0k}(c,x)$ has exactly $k$ zeros in $(-1,1)$, 
and even and odd according as $k$ is even and odd.

Alternatively 
$S_{0k}(c,x)$ is also a solution of the integral equation 
\beq\label{Integral equation to define S_0k}
\chi_k(c) S_{0k}(c,x) 
=\int_{-1}^1 dy \frac{\sin c(x-y)}{\pi(x-y)}
S_{0k}(c,y), 
\quad
\vert x\vert\le1. 
\eeq
The eigenvalues $\chi_k(c)$ are expressed by using 
a second set of solution $R_{0k}^{(1)}(c,x)$ for 
(\ref{differential equation}), 
called the radial prolate spheroidal wave functions, 
as 
\beq
\chi_k(c)=\frac{2c}{\pi}\left[R_{0k}^{(1)}(c,1)\right]^2.  
\eeq
$R_{0k}^{(1)}(c,x)$ differ from $S_{0k}(c,x)$ 
only by a real scale factor.

From Eq. (\ref{Integral equation to define S_0k}), 
we have 
\beq\label{Integral equation for Psi}
\chi_k(c) \Psi_{k}(c,t) 
=\int_{-T/2}^{T/2} dt' 
\frac{\sin \pi B(t-t')}
     {\pi(t-t')}
\Psi_{k}(c,t'),  
\eeq
and 
\beq\label{Integral equation for Phi}
\chi_k(c) \Phi_{k}(c,\Omega) 
=\int_{-\pi B}^{\pi B} d\Omega' 
\frac{\sin \frac{(\Omega-\Omega')T}{2}}
     {\pi(\Omega-\Omega')}
\Phi_{k}(c,\Omega').  
\eeq
The orthogonal condition of 
Eq. (\ref{Orthogonal condition for Psi}) 
can be derived by using 
Eq. (\ref{Integral equation for Psi}) 
and the orthonormal condition 
\beq\label{Normalization condition for Psi}
\int_{-\infty}^{\infty} dt
\Psi_k(c, t)\Psi_l(c, t)=\delta_{kl}. 
\eeq

Another important relation is 
\beq
2i^k R_{0k}^{(1)}(c,1) S_{0k}(c,x) 
=\int_{-1}^1 dy e^{icxy}
S_{0k}(c,y), 
\eeq
from which the Fourier transformation, Eq. 
(\ref{FT of prolate spheroidal functions}) can be derived. 
From Eqs. (\ref{FT of prolate spheroidal functions}), 
and (\ref{Integral equation for Phi}), we have 
\beq\label{FT of prolate spheroidal functions 2}
\chi_k(c)\Phi_k(c, \Omega)
=
\int_{-T/2}^{T/2} dt
\Psi_k(c, t) e^{i\Omega t}.
\eeq

\end{document}